\documentclass[floats,amsmath,amssymb,pra,aps,showpacs,superscriptaddress,twocolumn]{revtex4-1}

\pdfoutput=1
\usepackage{amsmath,amsfonts,amssymb,amsthm,graphics,graphicx,epsfig,bbm}
\usepackage[colorlinks=true,citecolor=blue,linkcolor=blue,urlcolor=blue]{hyperref}
\usepackage[usenames]{color}
\usepackage{graphicx}
\usepackage{subfigure}
\usepackage{amsmath}
\usepackage{epsfig}
\usepackage{dcolumn}
\usepackage{bm}
\usepackage{color}
\usepackage{epstopdf}
\usepackage{amssymb}
\usepackage{amstext}
\usepackage{latexsym}
\usepackage{hyperref}
\usepackage{amsfonts}
\usepackage{psfrag}
\usepackage{xcolor}
\usepackage[normalem]{ulem}
\usepackage{dsfont}
\usepackage{txfonts}
\usepackage{ifthen}

\usepackage{subfigure,float,morefloats}

\graphicspath{{./Images/}{./AppendixImages/}}

\newcommand{\ket}[1]{\ensuremath{\left\vert #1 \right\rangle}}

\newcommand{\mean}[1]{\left\langle #1 \right\rangle}

\newcommand{\phiA}{\ensuremath{\phi_\text{A}}}
\newcommand{\phiB}{\ensuremath{\phi_\text{B}}}

\newcommand{\an}[2]{\ensuremath{\hat{#1}^{\protect\phantom{\dagger}}_{#2}}}
\newcommand{\cn}[2]{\ensuremath{\hat{#1}^\dagger_{#2}}}
\newcommand{\nn}[2]{\ensuremath{\hat{n}^{#1}_{#2}}}

\newcommand{\expU}[1]{\ensuremath{e^{#1}}}
\newcommand{\abs}[1]{\left|#1\right|}

\newcommand{\replyA}[1]{\blue { #1}}

\newcommand{\be}[1]{\begin{eqnarray} \label{e#1}}
\newcommand{\beq}{\begin{eqnarray}}
\newcommand{\eeq}{\end{eqnarray}} 
\newcommand{\hide}[1]{}
 
\newcommand{\hrefl}[1]{\href{#1}{[link]}}

\newcommand{\blue}[1]{\textcolor{blue}{#1}}

\newcommand{\subfigimg}[3][,]{%
		\setbox1=\hbox{\includegraphics[#1]{#3}}
		\leavevmode\rlap{\usebox1}
		\rlap{\hspace*{2pt}\raisebox{\dimexpr\ht1-0.5\baselineskip}{{\bfseries \large\textsf{#2}}}}
		\phantom{\usebox1}
}

\newcommand\numberthis{\addtocounter{equation}{1}\tag{\theequation}}

\newcommand{\idg}[1]{{\bfseries #1)}}

\begin{document}

\title{Nonclassical states in strongly correlated bosonic ring ladders}

\author{Nicolas Victorin}
\affiliation{Univ. Grenoble-Alpes, CNRS, LPMMC, 38000 Grenoble, France}
\author{Tobias Haug}
\affiliation{Centre for Quantum Technologies, National University of Singapore,
3 Science Drive 2, Singapore 117543, Singapore}
\author{Leong-Chuan Kwek}
\affiliation{Centre for Quantum Technologies, National University of Singapore,
	3 Science Drive 2, Singapore 117543, Singapore}
\affiliation{MajuLab, CNRS-UNS-NUS-NTU International Joint Research Unit, UMI 3654, Singapore}
\affiliation{Institute of Advanced Studies, Nanyang Technological University,
	60 Nanyang View, Singapore 639673, Singapore}
\affiliation{National Institute of Education, Nanyang Technological University,
	1 Nanyang Walk, Singapore 637616, Singapore}
\author{Luigi Amico}
\affiliation{Centre for Quantum Technologies, National University of Singapore,
3 Science Drive 2, Singapore 117543, Singapore}
\affiliation{MajuLab, CNRS-UNS-NUS-NTU International Joint Research Unit, UMI 3654, Singapore}
\affiliation{Dipartimento di Fisica e Astronomia, Via S. Sofia 64, 95127 Catania, Italy}
\affiliation{CNR-MATIS-IMM \&   INFN-Sezione di Catania, Via S. Sofia 64, 95127 Catania, Italy}
\affiliation{LANEF {\it 'Chaire d'excellence'}, Universit\`e Grenoble-Alpes \& CNRS, F-38000 Grenoble, France}
\author{Anna Minguzzi}
\affiliation{University Grenoble-Alpes, CNRS, LPMMC, 38000 Grenoble, France}

\begin{abstract}
  We study the ground state of a bosonic ring ladder under a gauge flux in the vortex phase, corresponding to the case where the single-particle dispersion relation has two degenerate minima. By combining exact diagonalization and an approximate fermionization approach we show that the ground state of the system evolves from a fragmented state of two single-particle states at weak interparticle interactions to a fragmented state of two Fermi seas at large interactions. Fragmentation is inferred from the study of the eigenvalues of the reduced single-particle density matrix as well as from the calculation of the fidelity of the states. We characterize these nonclassical states by the momentum distribution, the chiral currents and the current-current correlations.
\end{abstract}

\maketitle

\section{Introduction}

Recent progress in ultracold lattice bosons allow us to study strongly correlated many-body systems with enhanced tunability \cite{bloch2008many}.
One of the problems that have attracted interest in the community is a system of ladders pierced by an external magnetic flux.
Indeed, such system provide an exciting playground to address relevant aspects of many-body correlated matter such as the Meissner effect in type II superconductors and quantum Hall effect in semiconductors\cite{kardar1986josephson,orignac2001meissner,petrescu2015chiral}. 
Flux ladders have been already experimentally realized in a linear geometry with state of the art ultra-cold atoms quantum technology\cite{atala2014observation,mancini2015observation,stuhl2015visualizing,an2016direct,livi2016synthetic} .
Depending on the system parameters, interaction and effective magnetic field,  the system enters different physical regimes characterized by distinctive current patterns\cite{petrescu2013bosonic,tokuno2014ground,wei2014theory,di2015meissner,orignac2016incommensurate,greschner2015spontaneous,piraud2015vortex,di2015meissner}. 

Here we focus on the new avenue of research  studying ultracold atoms loaded in closed geometries\cite{dalibard2011colloquium,wright2013driving,Ramanathan2011,Ryu2013,eckel2014hysteresis,yakimenko2015,eckel2014hysteresis,hallwood2006macroscopic,solenov2010metastable,schenke2011nonadiabatic,amico2014superfluid,aghamalyan2015coherent,aghamalyan2016atomtronic,Mathey_Mathey2016,haug2017aharonov,haug2018readout,haug2018andreev}. Atomtronics, in particular, promise to substantially enlarge the scope of cold atoms quantum simulators, and put the basis for new quantum devices and sensors \cite{seaman2007atomtronics,amico2005quantum,dumke2016roadmap,Amico_NJP}. Here, we consider a specific atomtronic network made of two coupled rings, which provide an analogical quantum simulator of bosonic flux ladders (see also \cite{aghamalyan2013effective}).
The ring geometry allows to naturally study the pattern of current flows through the ladder, as well as minimizes the effects of boundary conditions in the quantum simulator. In coupled ring lattices,  currents, time-of-flight images and spiral interferograms have been studied in various regimes of interactions, from mean-field approaches for large fillings and  weak interactions \cite{victorin2018bosonic}  to exact diagonalization methods for small fillings and arbitrary interactions \cite{haug2018mesoscopic}.

A bosonic system is in a single  Bose-Einstein condensate if its single-particle density matrix has one  macroscopic eigenvalue (i.e order of the number of particle) \cite{PenroseOnsager}. If the single-particle density matrix has more than one macroscopically occupied  eigenvalue, then the state is named fragmented \cite{Noziere,MuellerFrag}.
Nozi{\`e}res and Saint James \cite{Noziere} demonstrated that no fragmentation can take place in a homogeneous Bose gas with repulsive interactions.
For dispersion relations with degenerate minima, instead, fragmented states may emerge \cite{Spin1Frag,RotFrag,kawasaki2017finite}.

Such a type of dispersion relation occurs in the  vortex phase of double ring lattices, displaying  in particular a two-minima structure. 
Here, we investigate the nature of the system's ground state at arbitrary interactions.  The mean-field approach assumes a coherent state, made of superposition of single-particle  occupancies of each minimum. However,  it has been shown that the ground state at small lattice fillings and weak interactions is indeed a fragmented state constructed with  single-particle momentum states \cite{kolovsky2017bogoliubov}. This result can be related  the studies of spin-orbit coupled systems, which share the same type of Hamiltonian as bosonic flux ladders, and where the fragmentation was also observed with an ab-initio numerical study~\cite{kawasaki2017finite}.  In this work, we explore the fate of the fragmented state at increasing interactions. In particular, we show that strong repulsive interactions destroy the fragmented single-particle state, and  give rise to a novel type of nonclassical state, which can be described as a fragmentation of two Fermi spheres. The ground state crossover is  analyzed by studying  the correlation functions as implied in the the one-body density matrix and the specific changes in the configuration of the currents flowing in the ladder.

The paper is outlined as follows. In Sec.\ref{model}, we introduce the model Hamiltonian and discuss the different entanglement properties of the ground states  in the different physical regimes of the ladder; in addition we sketch the analytical methods (an approximate fermionization scheme) that we employ to study the different system's observables we refer to.
In the Sec.\ref{results}, we present the results obtained with the analytical methods and compare them with the exact diagonalization; our findings are corroborated by the study of the configuration of currents flowing in the ring ladder. Finally Sect. \ref{conclusions}, is devoted to a summary and concluding remarks.
\section{Model system and methods}
\label{model}
We consider two ring lattices $A$ and $B$, occupied by bosons at zero temperature, with the same number of sites in each ring  and tunnel coupled via the rungs.
The total Hamiltonian of the system reads $\mathcal{H}=\mathcal{H_\text{A}}+\mathcal{H_\text{B}}+\mathcal{H_\text{I}}$, where 
\begin{align*}
\mathcal{H_\text{A}}={}&\sum_{m=1}^{L}\left(-J\expU{i2\pi\phiA}\cn{a}{m}\an{a}{m+1}+\text{h.c.}\right)+\frac{U}{2}\nn{A}{m}(\nn{A}{m}-1)\\
\mathcal{H_\text{B}}={}&\sum_{m=1}^{L}\left(-J\expU{i2\pi\phiB}\cn{b}{m}\an{b}{m+1}+\text{h.c.}\right)+\frac{U}{2}\nn{B}{m}(\nn{B}{m}-1)\label{Hamilton}\numberthis\\
\mathcal{H_\text{I}}={}&-K\sum_{m=1}^{L}\left(\cn{a}{m}\an{b}{m}+\text{h.c.}\right) \; ,
\end{align*}
where $J$ is the intra-ring tunnel energy, $K$ is the inter-ring tunnel energy, $U$ is the on-site interaction energy and $\phi_\text{A}$, $\phi_\text{B}$ are the artificial gauge fields applied to each ring, which we assume to be separately tunable.  In the following we will restrict for simplicity  \footnote{The consequences of other choices for $\phi_A$ and $\phi_B$ are discussed in \cite{victorin2018bosonic}.} to the case $\phi_\text{A}=\phi$, $\phi_\text{B}=-\phi$, corresponding to the case where counter-propagating currents are driven by the gauge fields on the two rings.

In order to characterize the ground state of the system we use various observables: the momentum distribution in ring A is defined as $n_k^A=\langle a_k^\dag a_k\rangle$, with ${a_k=\frac{1}{\sqrt{L}}\sum_{m=1}^{L}a_m\expU{i2\pi m/L}}$, and similarly we have $n_k^B=\langle b_k^\dag b_k\rangle$ for ring B.
The current operator at position $m$ along  the  same ring is defined as ${j_{a,m}^{\parallel}=-iJ  (a_m^\dag a_{m+1} -\text{h.c.})}$.
We define the chiral current in the ladder as $j_\text{c}=\frac{1}{L}\sum_m (j_{a,m}^{\parallel} -j_{b,m}^{\parallel})$.
The  inter-ring current operator  at site $m$ is ${j_m^\bot=-iK (a_m^\dag b_m -\text{h.c.})}$. 
Correspondingly, we define the current-current correlation of the current between the rings $\mean{j_0^\bot j_m^\bot}$ and the current inside the ring $\mean{j_0^\parallel j_m^\parallel}$. 
We also investigate the density-density correlations in the leg $\Delta n_m=\langle n_0 n_m \rangle- \langle n_0 \rangle\langle n_m \rangle$ and between legs $\Delta n_m^\bot=\langle n_0^A n_m^B \rangle- \langle n_0^A \rangle\langle n_m^B \rangle$.

\label{ansatz}
By Fourier transforming, the Hamiltonian (\ref{Hamilton}) is readily expressed in momentum space according to 
\begin{align*}
  \label{Ladderbasis}
\mathcal{H_\text{A}}={}&\sum_{k}-2J\cos(k+2\pi\phi)\nn{A}{k}+\frac{U}{2L}\sum_{k,k',q}\cn{a}{k+q}\cn{a}{k'-q}\an{a}{k'}\an{a}{k}\\
\mathcal{H_\text{B}}={}&\sum_{k}-2J\cos(k-2\pi\phi)\nn{B}{k}+\frac{U}{2L}\sum_{k,k',q}\cn{b}{k+q}\cn{b}{k'-q}\an{b}{k'}\an{b}{k}\numberthis\\
\mathcal{H_\text{I}}={}&-K\sum_{k}\left(\cn{a}{k}\an{b}{k}+\text{h.c.}\right) \; .
\end{align*}
It is useful to transform the above Hamiltonian to a new basis, which is diagonal in absence of  interactions. We call this the ``diagonal" basis. By using the transformation
\begin{align}
\begin{pmatrix}
a_{k}\\
b_{k}
\end{pmatrix}
=
\begin{pmatrix}
v_k & u_k\\
-u_k & v_k
\end{pmatrix}
\begin{pmatrix}
\alpha_k\\
\beta_k
\end{pmatrix},
\end{align}
where $u_k$ and $ v_k$ are given by 
\begin{eqnarray}
v_k = \sqrt{\frac{1}{2}(1+\frac{\sin(2 \pi \phi)\sin(k)}{\sqrt{(K/2J)^2+\!\sin^2(2 \pi \phi)\sin^2(k))}})}\\
u_k = \sqrt{\frac{1}{2}(1-\frac{\sin(2 \pi \phi)\sin(k)}{\sqrt{(K/2J)^2+\!\sin^2(2 \pi \phi)\sin^2(k))}})}
\end{eqnarray}
the non-interacting part of the Hamiltonian becomes
\begin{eqnarray}  \label{HamSym}
\hat{H}_0 = \sum_k \alpha_k^{\dagger}\alpha_kE_+(k) + \beta_k^{\dagger}\beta_k E_-(k),
\end{eqnarray}
with  ${E_\pm=-2J\cos(k)\cos(2\pi\phi)\pm\sqrt{K^2+4J^2\sin^2(k)\sin^2(2\pi\phi)}}$. In the vortex phase the lowest branch  $E_-(k)$ of the dispersion relation has two degenerate minima at  ${k_{1,2}=\pm\arcsin \sqrt{\sin^2(2\pi\phi)-\frac{K^2}{(2J\tan(2\pi\phi))^2}}}$.
It will be useful to study the momentum distributions in the diagonal basis, ie  the one of the lower branch $n_\beta(k)=\langle \beta_k^\dag \beta_k\rangle$ and the one of the upper branch $n_\alpha(k)=\langle \alpha_k^\dag \alpha_k\rangle$

In the vortex phase, for very small ring-ring coupling $K$ and small, non-zero interaction strength $U$, the ground state is fragmented, {\it i.e.} displays macroscopic occupation of the two single-particle momentum states $k_{1,2}$ as described by the  the Ansatz \cite{kolovsky2017bogoliubov}
\begin{equation}
\ket{\Psi_0^{(sp)}}=\cn{\beta}{k=k_1}{}^{N/2}\cn{\beta}{k=k_2}{}^{N/2}\ket{0}\;.
\label{frag-sp}
\end{equation}
Notice that the ground state is built only with the field operators $\beta$  associated to the lowest branch of the dispersion relation, and the problem has been mapped to an effectively one-dimensional one. In the presence of interactions, the effective one-dimensional Hamiltonian restricted to the lowest branch reads 
\begin{align}
H=&\sum_kE_-(k)\beta_k^{\dagger}\beta_k\nonumber\\&+\frac{U}{2N_s}\sum_{q,k,r}\kappa(k-q,r+q,k,r)\beta^{\dagger}_{k-q}\beta^{\dagger}_{r+q}\beta_{k}\beta_r
\end{align}
where the kernel $\kappa(k-q,r+q,k,r)=u_{k-q}u_{r+q}u_{k}u_{r}+v_{k-q}v_{r+q}v_{k}v_{r}$ is an effective interaction potential in momentum space, which has some involved momentum structure. However, if the ratio  $K/J$   is small, the parameters $u_k$ and $v_k$ can be approximated as  constants for wavevector $k$ close to $k_1$ and $k_2$. In this case, for the sake of finding the ground state, the Hamiltonian is equivalent to the one of a one-dimensional Bose gas with contact interactions with a single-particle dispersion $E_-(k)$.

At increasing interaction strength, clearly the fragmented single-particle state Ansatz (\ref{frag-sp}) is  not expected to describe well the ground state state of the system, since repulsive interactions give rise to a spread in momentum occupancy. In the regime $U\to \infty$ of very strong repulsions, we predict an effective fermionization of the ground state, ie two particles cannot occupy the same momentum state and we propose the following fragmented Fermi-sea Ansatz:
\begin{equation}
\ket{\Psi_0^{(Fs)}}=\prod_{-k_F+k_1<k<k_F+k_1} \cn{c}{k}{} \prod_{-k_F+k_2<k<k_F+k_2} \cn{c}{k}{}\ket{0}\;.
\label{frag-Fs}
\end{equation}
where $k_F$ is the Fermi wavevector corresponding to $N/2$ particles and $\cn{c}{k}{}$ the fermionic creation operator for the lower band of the non-interacting Hamiltonian (e.g. $\left\{\cn{c}{k}{},\cn{c}{k'}{}\right\}=\delta_{k,k'}$, where $\left\{.\right\}$ is the anti-commutator). This  Ansatz is only valid as long as the Fermi energy is smaller than the energy of the upper band, i.e $E_\text{Fermi}<E_+(k)$.

\subsection{Fermionization approach}
\label{fermionization}

Using the fragmented Fermi sea ground state, in the $U=\infty$ limit we calculate the one-body density matrix and the momentum distribution of the gas using a mapping onto noninteracting fermions.
In detail, we Fourier transform the Ansatz into real space, then apply the Jordan-Wigner transformation \cite{Jordan1928}
\begin{equation}
\hat{b}^{\dagger}_i=c^{\dagger}_{i}\prod_{j=1}^{i-1}e^{-i\pi c^{\dagger}_jc_j} \hspace{1cm}
 \hat{b}_i=\prod_{j=1}^{i-1}e^{i\pi c^{\dagger}_jc_j}c_{i}
\end{equation}
Hard-core bosons and non-interacting fermions have the same spectrum in one dimension. Differences between the two appears in off-diagonal correlation functions, eg in the one-body density matrix. Using the relation between the one-body density matrix $\rho_{ij}$ and the one-particle Green function $G_{ij}=\langle b_i b_j^{\dagger}\rangle$
\begin{equation}
\rho_{ij}=\langle b_i^{\dagger}b_j\rangle= G_{ij}+\delta_{ij}(1-G_{ii}),
\end{equation}
we calculate the one-body density matrix using a method developed by Rigol and Muramatsu \cite{RigolHCB}. In particular the one-particle Green function can be expressed in term of our Ansatz and fermionic operators according to
\begin{align}
G_{ij}&=\langle \Psi^{(HCB)}_0|b_ib^{\dagger}_{j}|\Psi^{(HCB)}_0\rangle \nonumber\\
&=\langle \Psi^{(Fs)}_0 |\prod_{p=1}^{i-1}e^{i\pi c^{\dagger}_pc_p}c_ic_j^{\dagger}\prod_{s=1}^{j-1}e^{-i\pi c_s^{\dagger}c_s}|\Psi^{(Fs)}_0\rangle
\end{align}
where $|\Psi^{(HCB)}_0\rangle$ is the ground state for hard-core bosons.

\section{Results}
\label{results}

\subsection{Fidelity with respect to fragmented states}

In order to infer the nature of the ground state of the system  we calculate the fidelity of the ground state obtained by exact diagonalization of the many-body Hamiltonian with the Lanczos algorithm and we project it onto  
the two Ansatz states discussed in Sec.\ref{ansatz}, i.e. we take 
$F=\abs{\left\langle\Psi_\text{0}\ket{\psi_\text{GS}\right.}}^2$ , where  $\ket{\Psi_\text{0}}$  is either the single-particle fragmented state Eq.(\ref{frag-sp})  or the Fermi-sea fragmented state Eq.(\ref{frag-Fs}). Notice that since we use a real-space basis for the numerical diagonalization we perform first a Fourier transform of the Ansatz state onto real-space.

\begin{figure}[htbp]
	\centering
	\subfigimg[width=0.24\textwidth]{a}{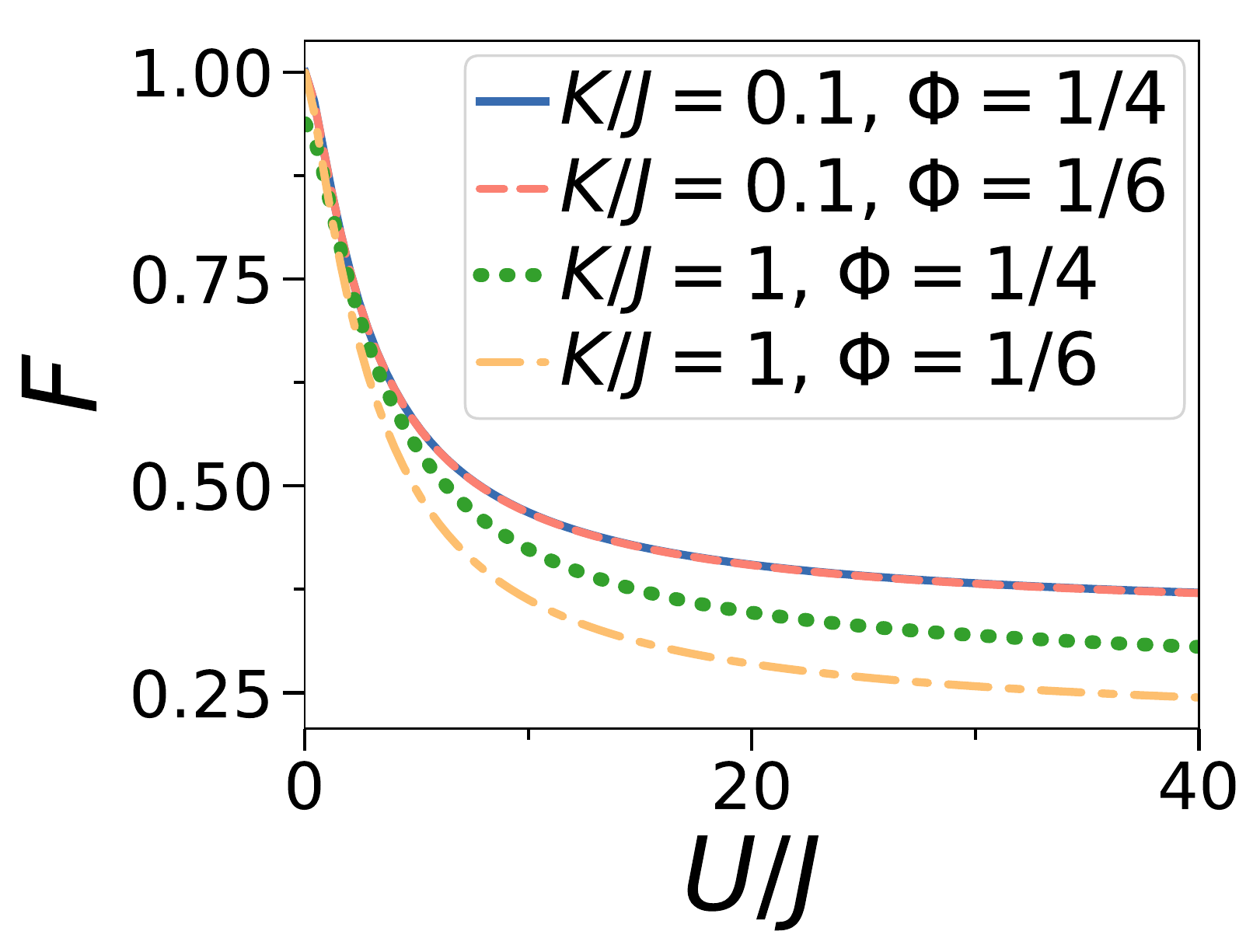}\hfill
	\subfigimg[width=0.24\textwidth]{b}{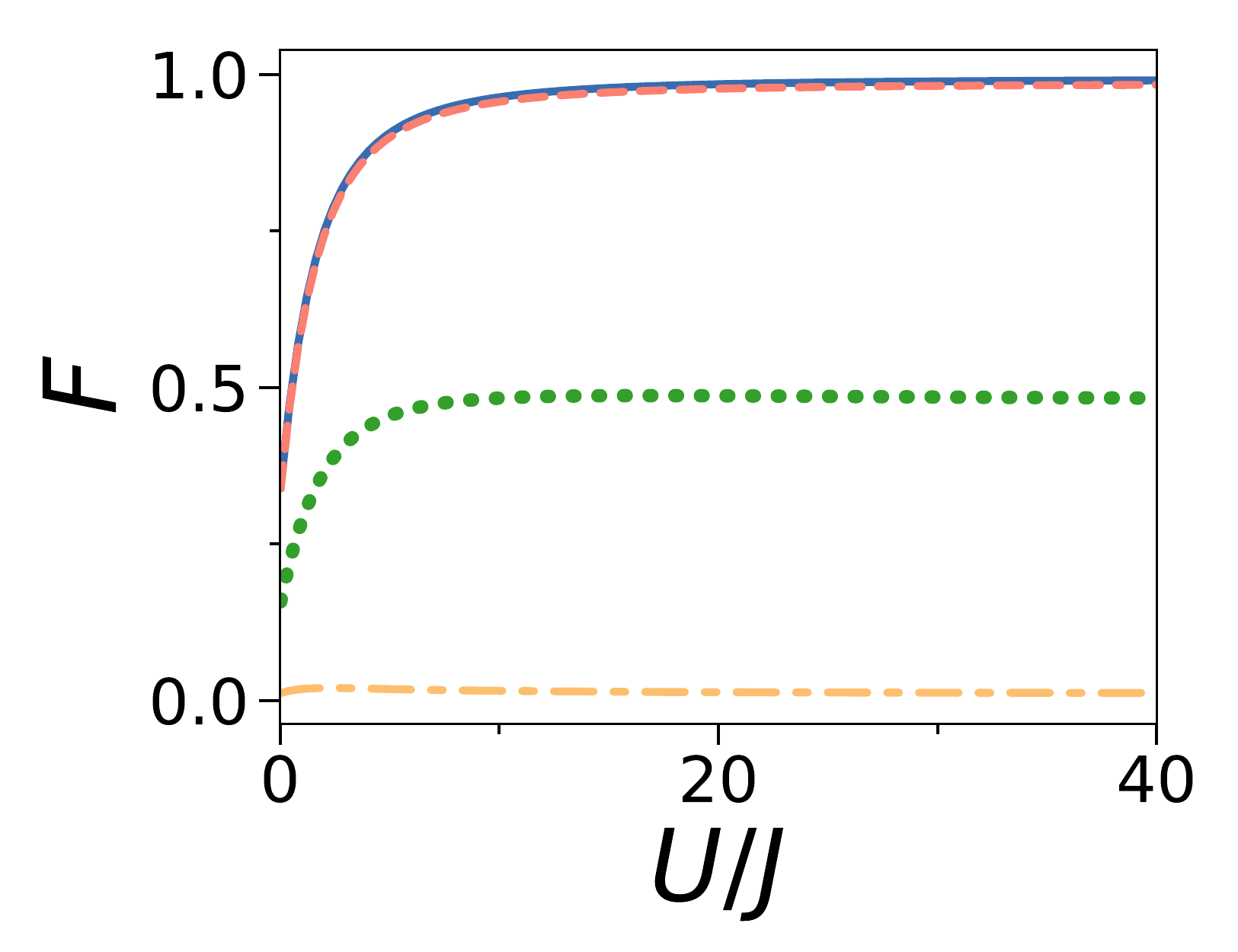}
	\caption{(Color online) Fidelity of the ground state obtained from numerical diagonalization as a function of interaction strength : \idg{a} with respect to the
single-particle fragmented state  $\ket{\Psi_0^{(sp)}}$
\idg{b} with respect to the fragmented Fermi sea  $\ket{\Psi_0^{(FS)}}$. The parameters used are $L=12$, $N=6$. The values of $K/J$ and $\Phi$ are indicated in the figure legend.}
	\label{fidelitySym}
\end{figure}

Our results are shown in  Fig.\ref{fidelitySym}. In the left panel we show that the fidelity with respect to the single-particle fragmented state decreases at increasing interactions. In the right panel, we show that the fidelity with respect to the fragmented Femi sea increases with interaction. For weak inter-ring coupling, we reach nearly unity fidelity for $U/J>10$ for any flux, thus confirming the validity of our Ansatz.

For strong inter-ring coupling, where the description of hard-core bosons breaks down, we find that the fidelity stays below one for any interaction  and its value  depends strongly on the choice of flux values. In this case there is no simple analytical description since bosons belonging to the lower band interact with a long-range interacting potential.

In the following, we provide an analysis of the observables characterizing the ground state of the system and identify the ones needed to infer the fragmented nature of the state in the vortex phase.

\subsection{Currents}
First, we show that the study of  chiral currents can be used  to identify univocally the vortex phase in parameter space, both in the interacting and non-interacting regime.

The Hamiltonian (\ref{Hamilton}) in absence of interactions  features the Meissner to vortex transition. At weak interactions, an additional biased ladder phase is found. At stronger interactions, chiral Mott insulating phase with Meissner like current and vortex Mott insulating phase are predicted \cite{petrescu2013bosonic,piraud2015vortex}. 

\begin{figure}[htbp]
	\centering
	\subfigimg[width=0.24\textwidth]{a}{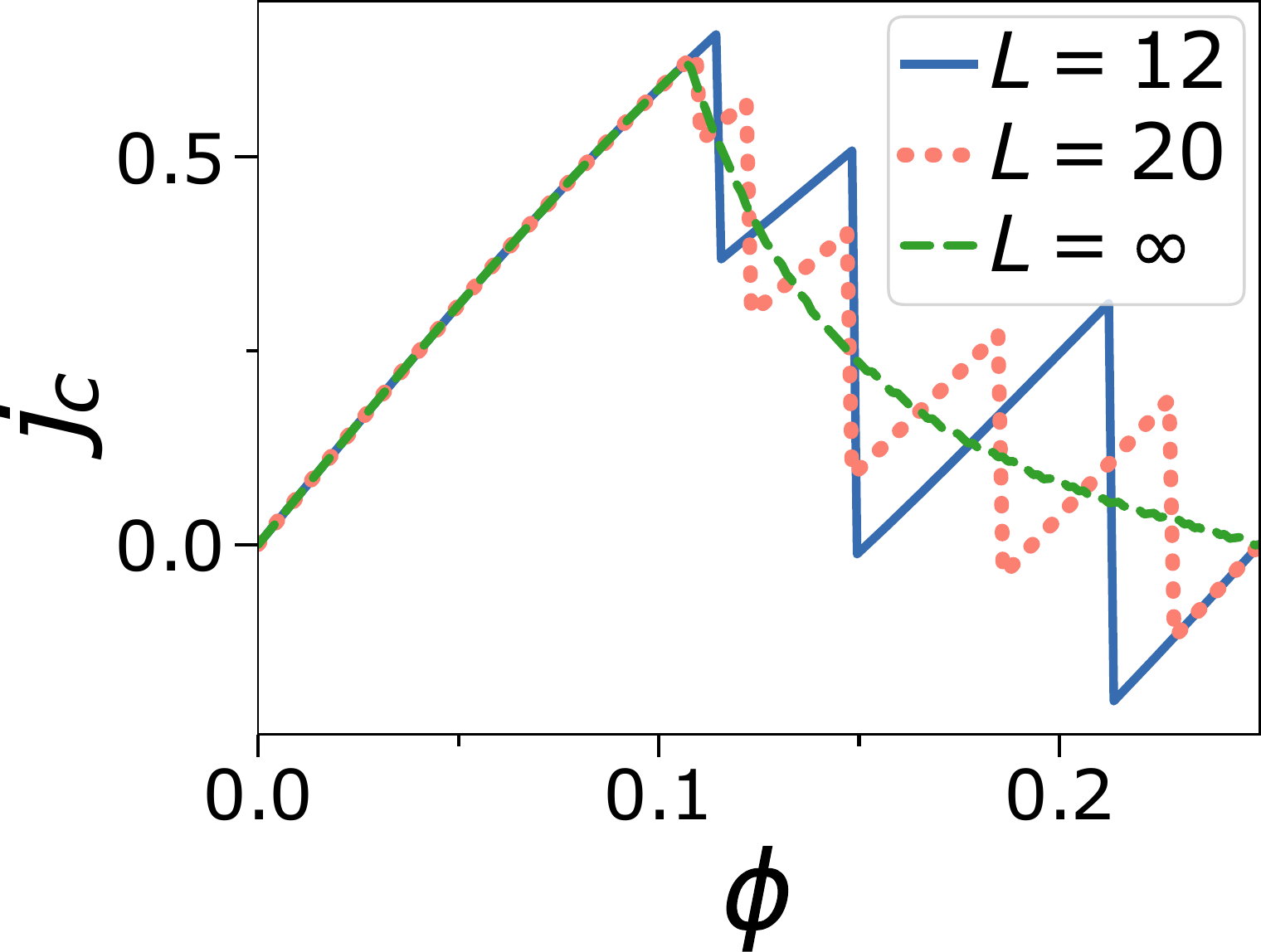}\hfill
	\subfigimg[width=0.24\textwidth]{b}{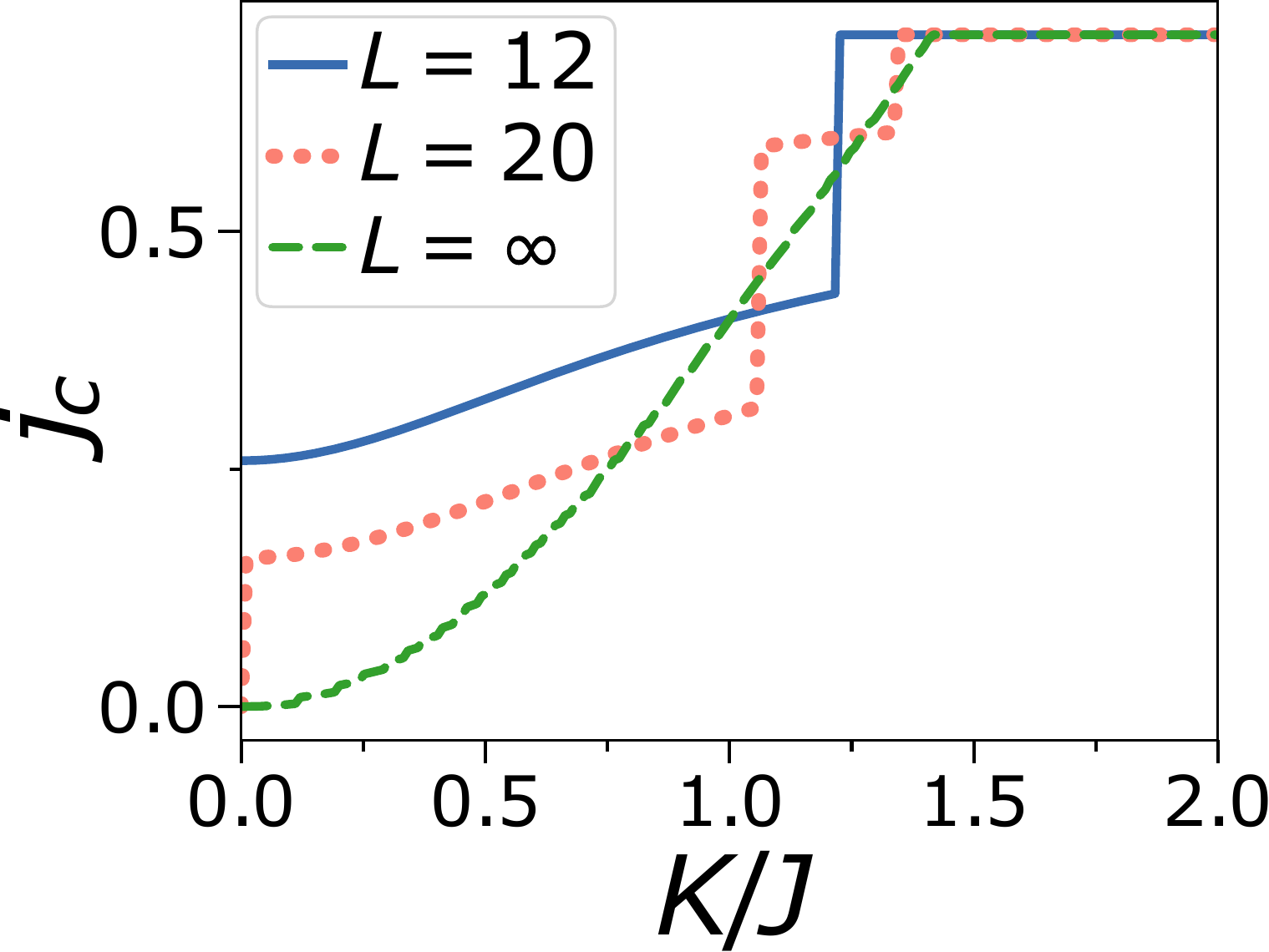}\\
	
	\caption{(Color online) \idg{a} Chiral current (in units of $J$) as a function of flux $\phi$ for $K/J=1$ for different ring lengths $L$. \idg{b} Chiral as a function of inter-ring coupling $K$ for $\phi=1/8$. 
	In both panels we have taken $U=0$ and half filling of the lattice.}
	\label{VMNA}
\end{figure}

In Fig.\ref{VMNA} we show the chiral current as a function of the gauge field $\phi$ in the noninteracting case. The Meissner phase has an increasing chiral current, whereas the current decreases in the vortex phase, where for finite-sized rings, the current acquires a step structure, each jump being associated to a integer change of the phase winding the formation of a vortex pair in the rings \cite{victorin2018bosonic}. In Fig.\ref{VMNA}b) we show the chiral current at increasing inter-ring tunneling $K/J$. We see that a change of behaviour occurs in the chiral currents in correspondence to the  transition from the vortex phase at low values of $K/J$ to the Meissner phase at large $K/J$ : a jump in the chiral current is found in the finite size-system while the chiral current is continuous with discontinuous derivative in the infinite-size limit.  We expect the transition to be of first order in analogy to the case of spin-orbit-coupled bosons \cite{Stringari-Li}.

\begin{figure}[htbp]
	\centering
	\subfigimg[width=0.24\textwidth]{a}{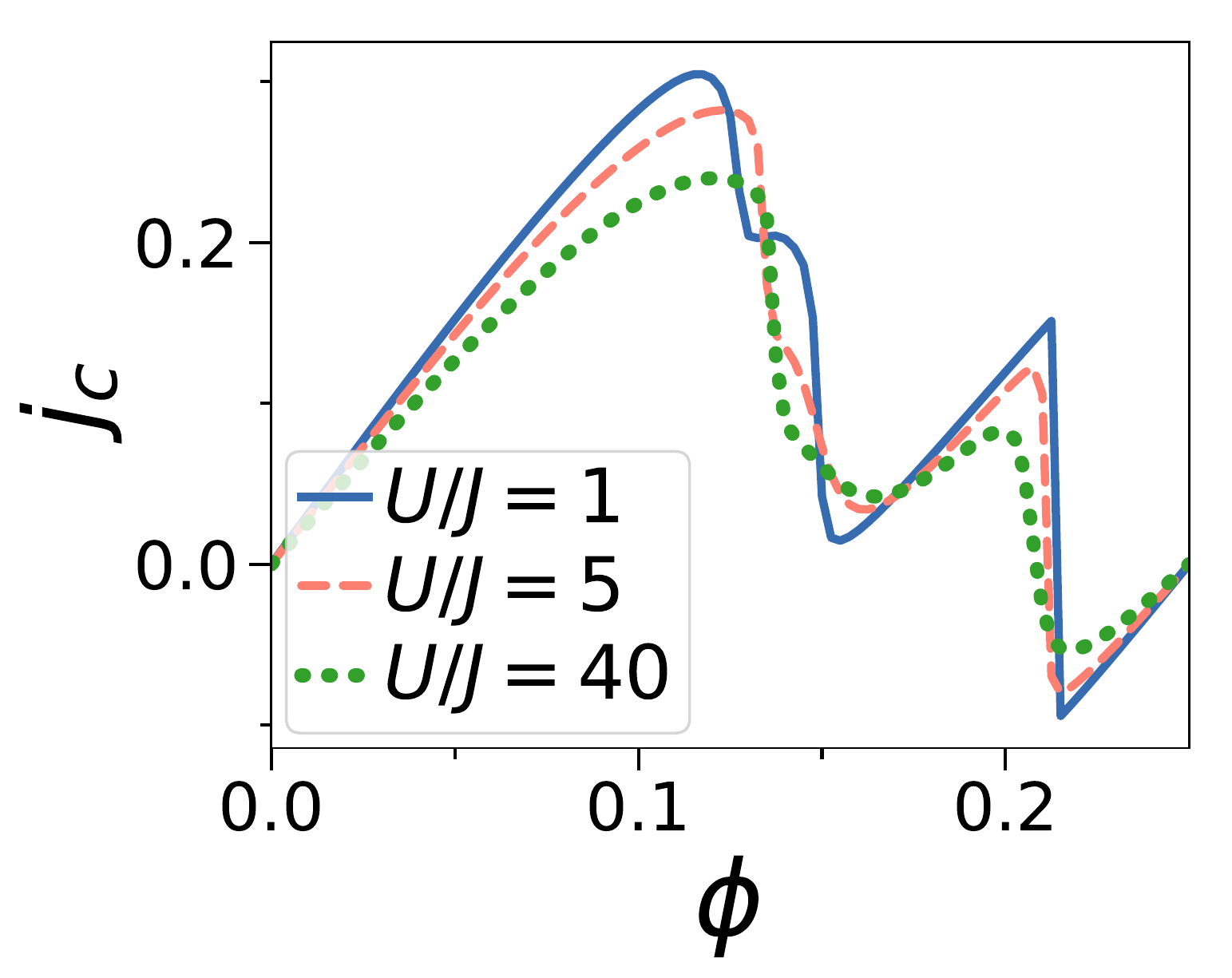}\hfill
	\subfigimg[width=0.24\textwidth]{b}{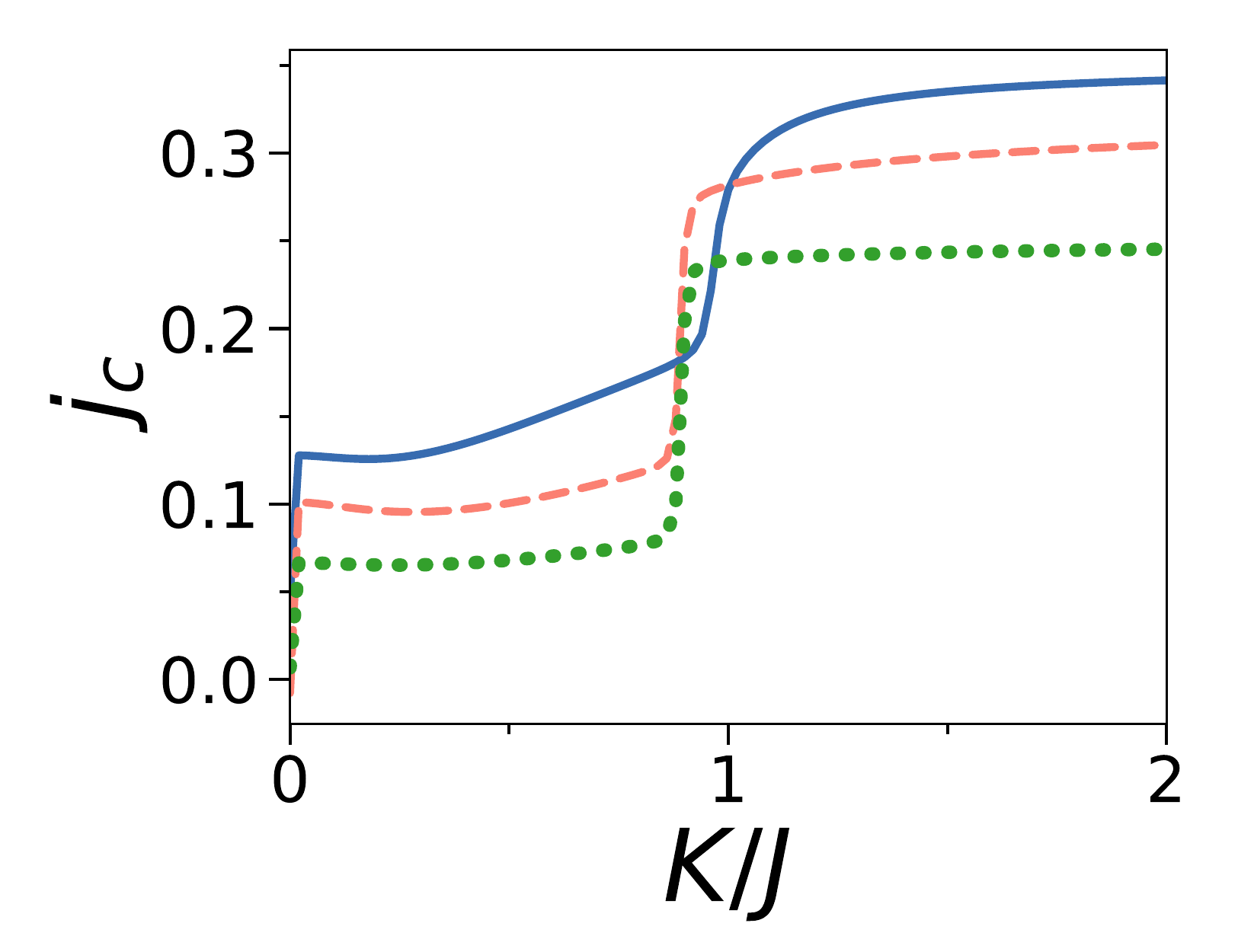}\\
	\caption{(Color online) \idg{a} Chiral current as a function of flux $\phi$ for $K/J=1$ and on-site interaction $U/J$ as indicated in the legend \idg{b} Chiral current as function of inter-ring coupling $K/J$ for $\phi=1/8$ and interaction strengh as indicated in the legend. The other parameters are ${L=12}$ and $N=6$ particles.}
	\label{VMNAInt}
\end{figure}

The chiral current for interacting system is shown in Fig.\ref{VMNAInt}.  We see that even though interactions smooth out the steps of the current and reduces the positions of the steps to lower values of $K/J$, overall it is still possible to infer the vortex phase as the regime where chiral current has a decreasing and oscillating behaviour as a function of the  flux $\phi$. Similarly to the non-interacting case, the transition from vortex to Meissner phase is visible by studying the dependence of chiral current on inter-ring coupling $K/J$.


\subsection{Current-current correlations} 
Another way to identify the vortex phase is the study of current-current correlations ~\cite{ChaMin2011}.

\begin{figure}[htbp]
	\centering
	\subfigimg[width=0.35\textwidth]{a}{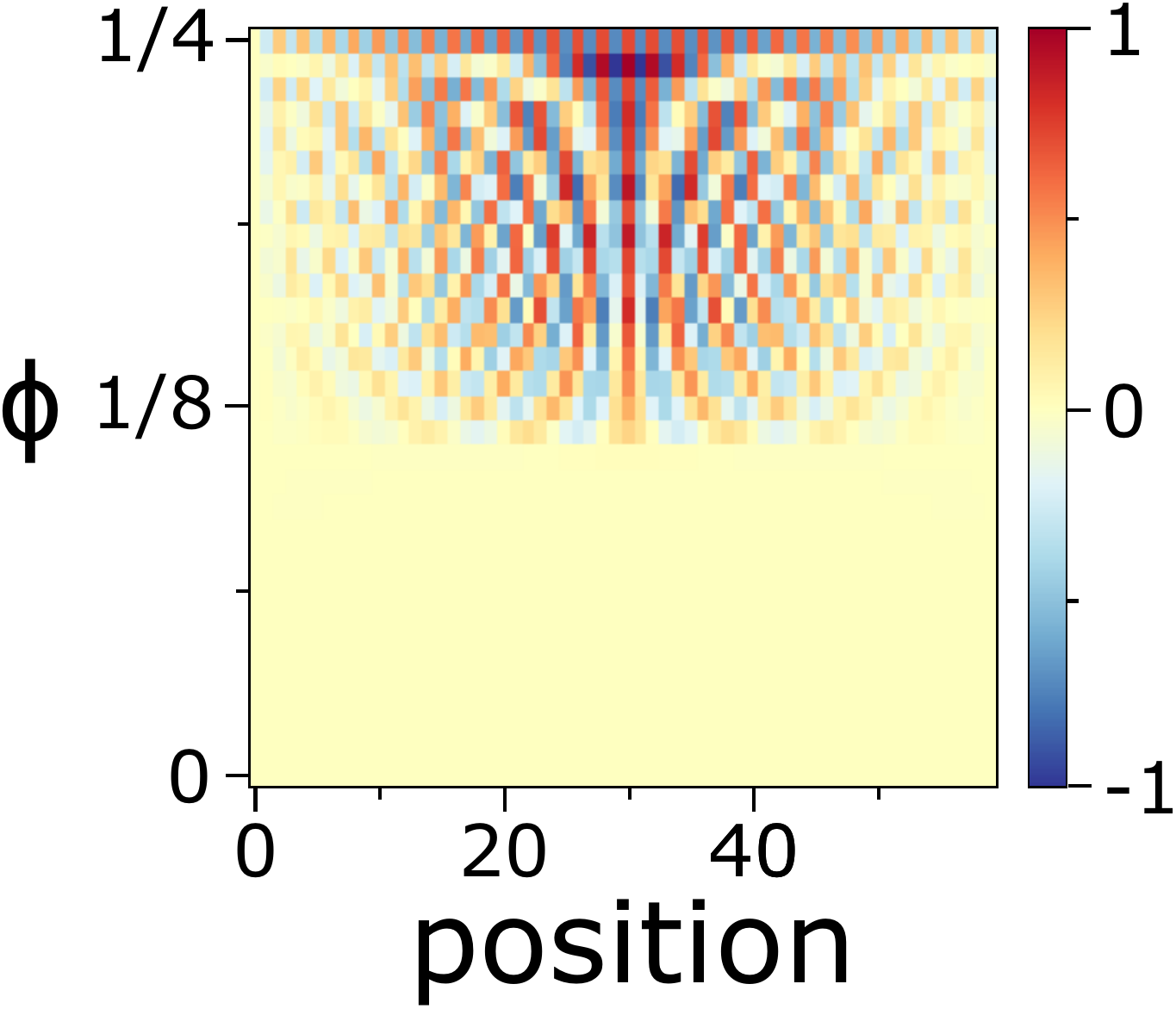}\hfill
	\caption{(Color online) Inter-ring current-current correlation $\mean{j_0^\bot j_x^\bot}$ in units of $J$ at varying flux $\Phi$ and lattice position for $L=60$, $N_\text{p}=2$, ${U/J=1}$ and ${K/J=1}$.}
	\label{currcurr}
\end{figure}

For linear ladders the ground state in the vortex phase is characterized by a vortex structure along the ladder,  a modulation of the density along the legs, and a modulation of the current between rungs. In the case of coupled lattice rings, corresponding to  periodic boundary conditions, these features are not visible since the ground state displays rotational invariance. In this case as the vortex phase sets in, the mentioned features are encoded in the correlation functions. In particular, the current-current correlation $\mean{j_0^{\perp}j_x^{\perp}}$, where $j_x^{\perp}$ is the current operator for the current between the rings, show a clear vortex structure in Fig.\ref{currcurr}.

\subsection{Density-density correlations}

Next, we show that  density-density correlations of the ground state may be used to infer the onset of strong correlations and fermionized regime.

\begin{figure}[htbp]
	\centering
	\subfigimg[width=0.24\textwidth]{a}{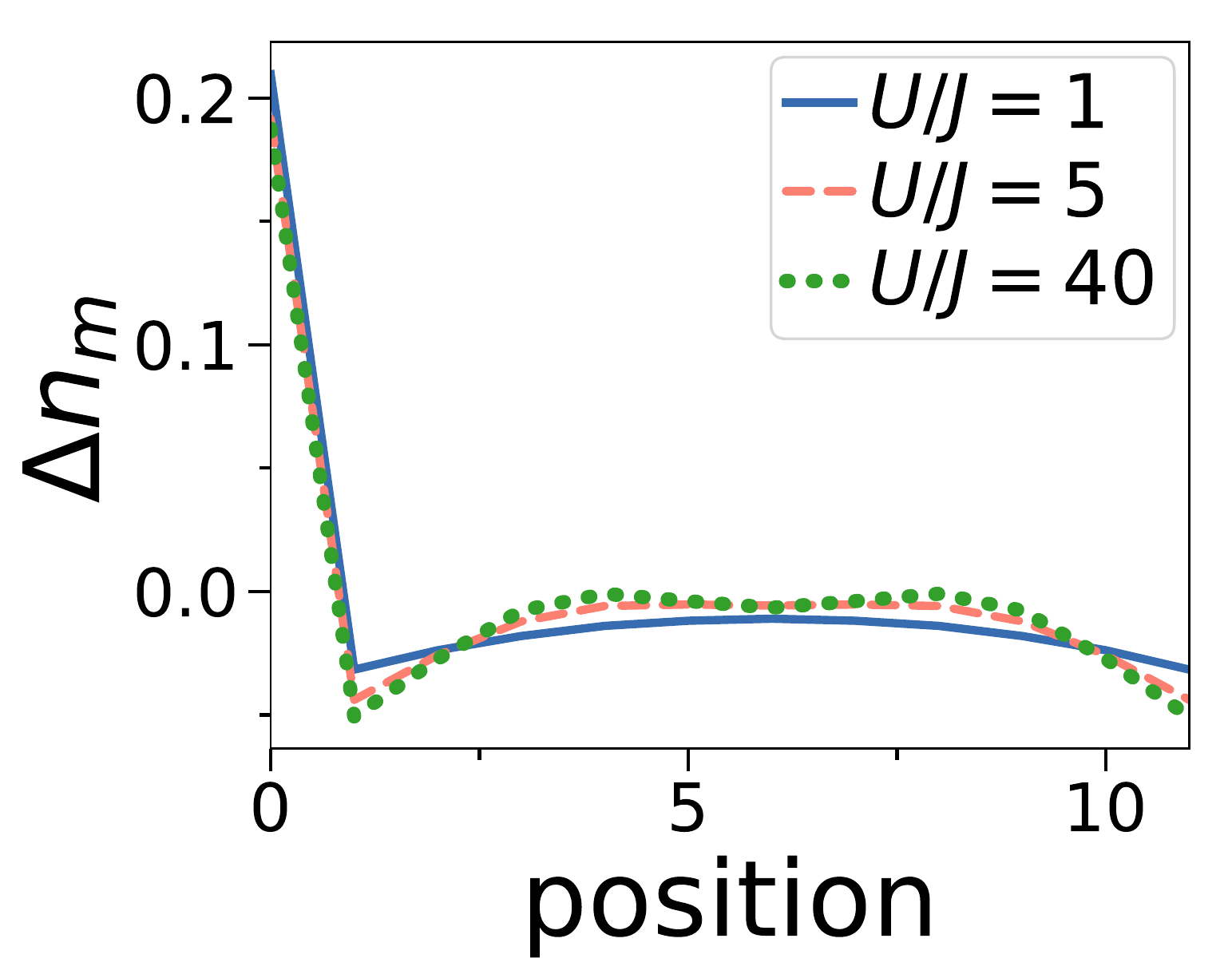}\hfill
	\subfigimg[width=0.24\textwidth]{b}{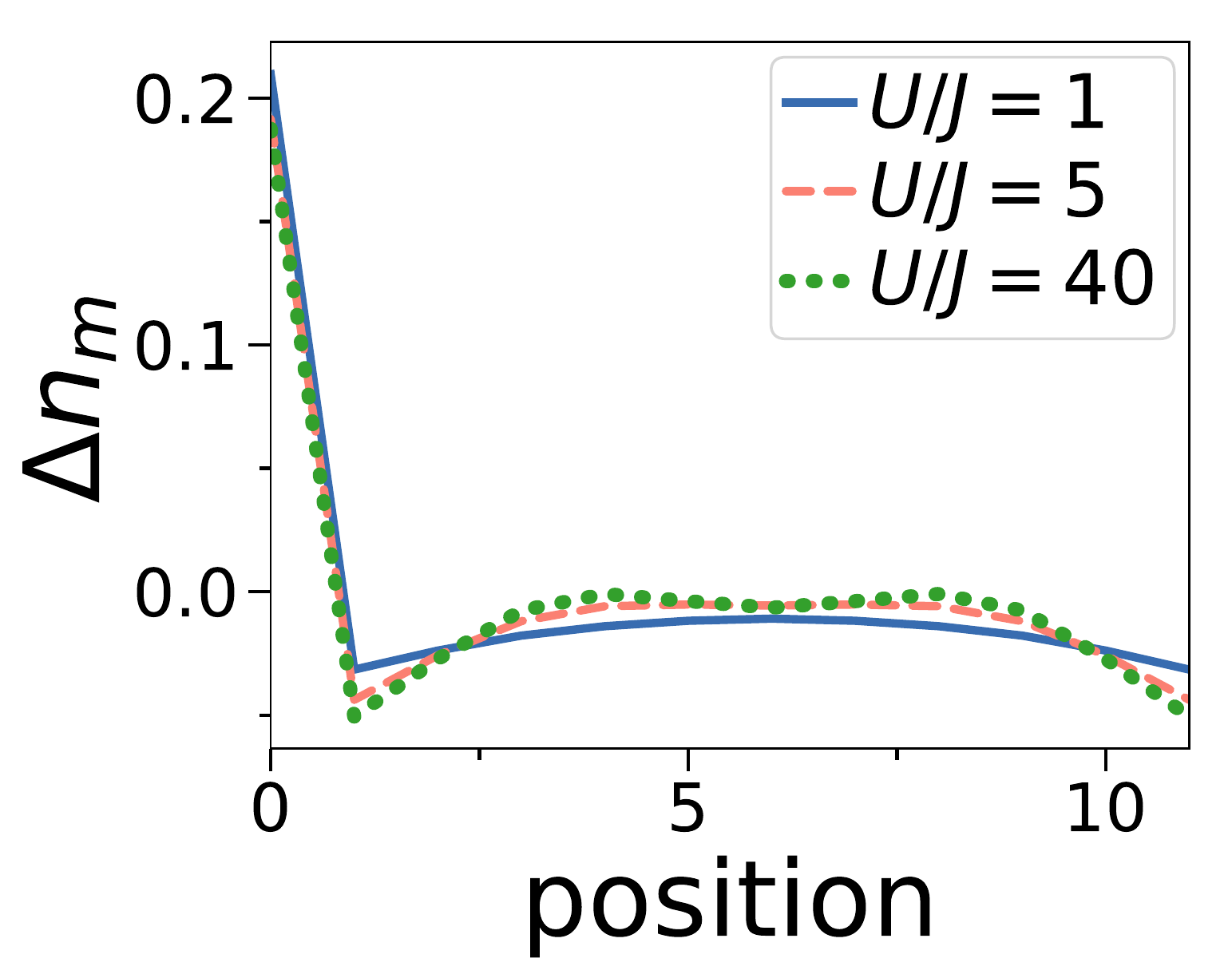}\\
	\subfigimg[width=0.24\textwidth]{c}{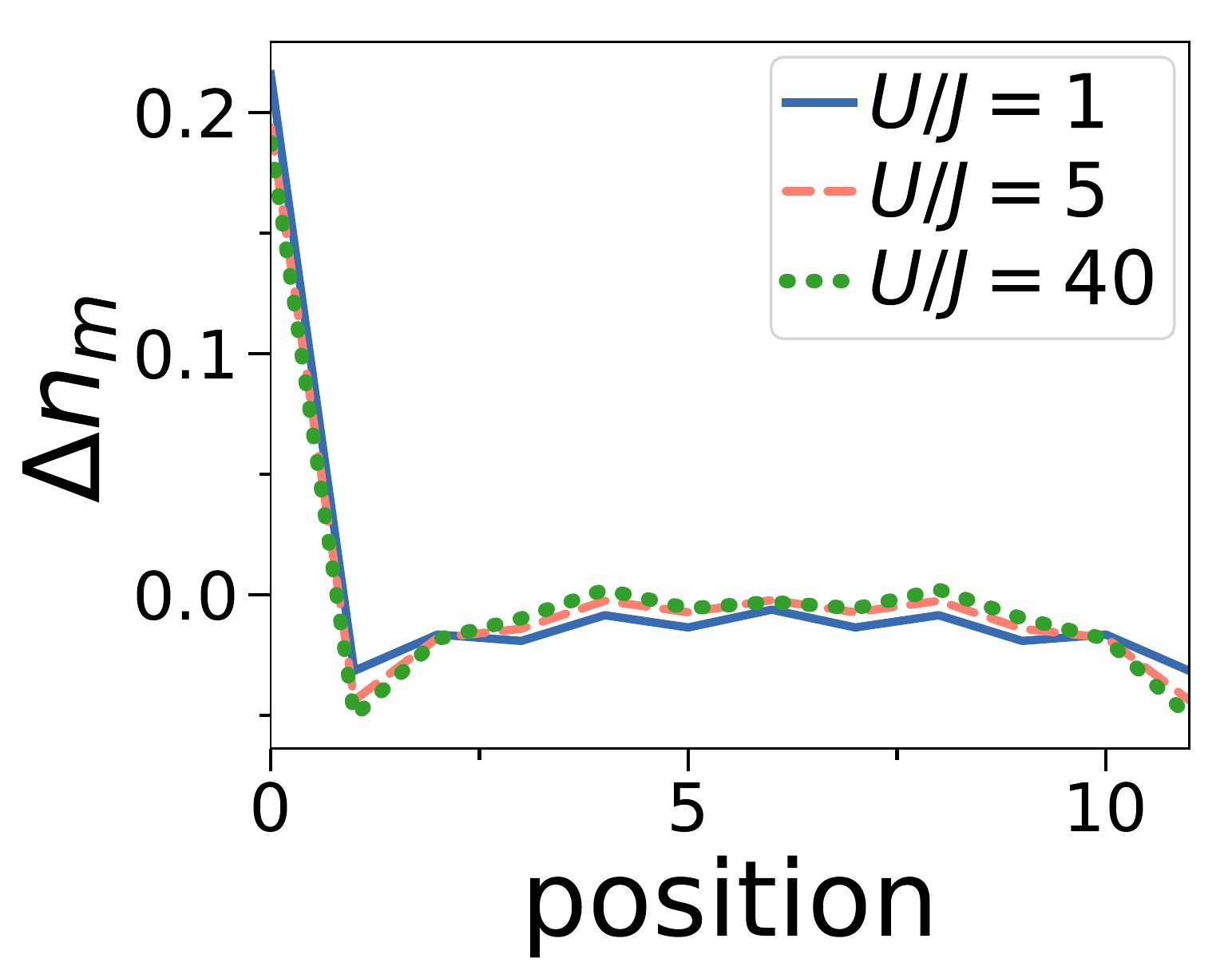}\hfill
	\subfigimg[width=0.24\textwidth]{d}{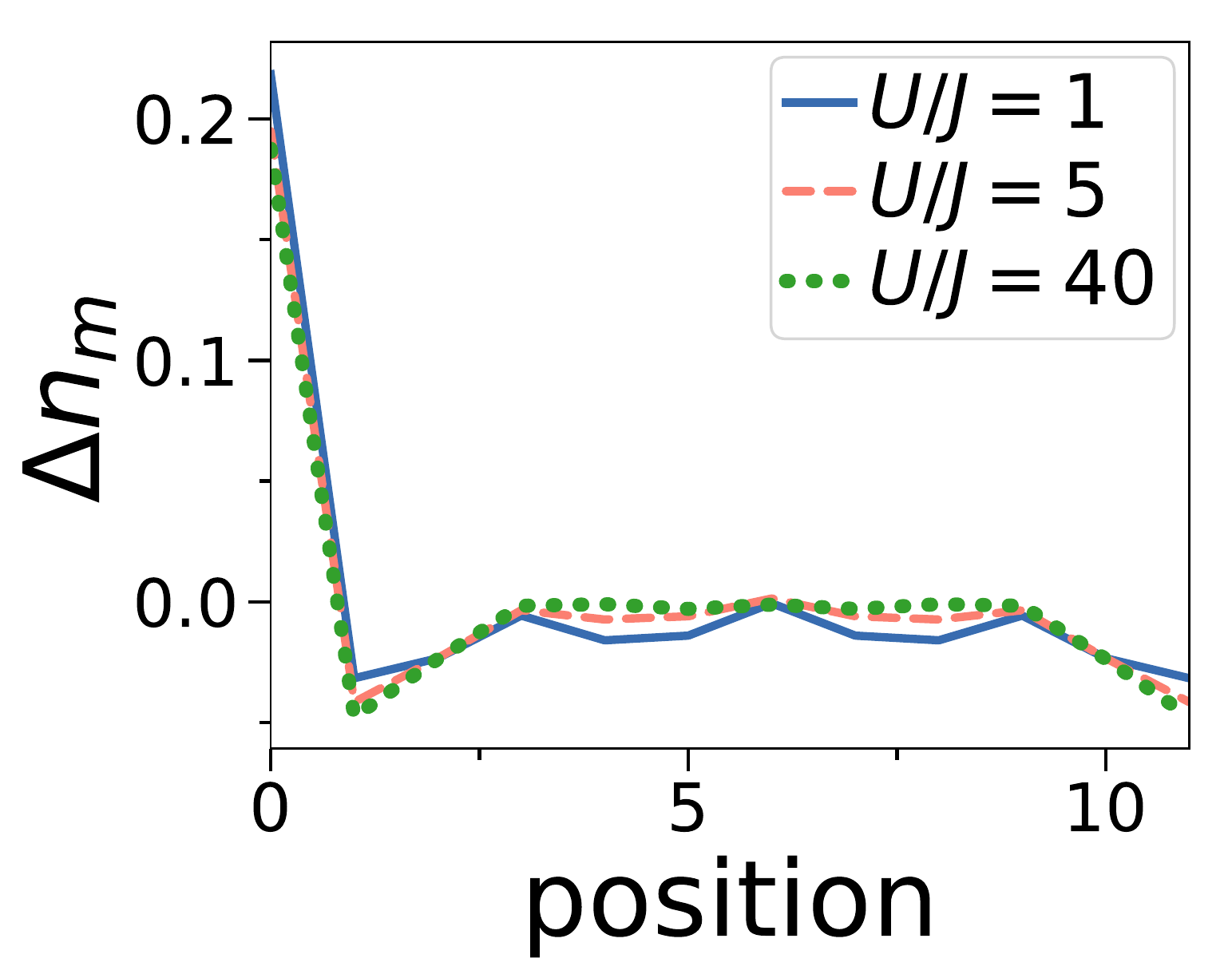}\\
	\caption{Density-density correlation function  $\Delta n_{m}$ in units of $J$ as a function of the position along the lattice.  \idg{a} $K/J=0.1$, $\phi=1/4$ \idg{b} $K/J=0.1$, $\phi=1/6$ \idg{c} $K/J=1$, $\phi=1/4$ \idg{d} $K/J=1$, $\phi=1/6$. The other parameters are $N=6$ and ${L=12}$.}
	\label{densdens}
\end{figure}

A hallmark of fermionization is the presence of Friedel-like oscillations, characterized by wavevevector $2k_F$, with $k_F$ the Fermi wavevector. These are found e.g.  in the  density-density correlation function  $\Delta n_{m}$, shown in  Fig.\ref{densdens} along one chosen ring (same results are found for the other ring). For small inter-ring coupling, we observe the build-up of Friedel-like oscillations at increasing interaction, with wavelength corresponding to four times the lattice spacing, corresponding to  the chosen average  lattice filling in each ring. For large values of $K/J$, the system is not any more quasi-one-dimensional. In this case we find that  the wavelength of Friedel oscillations changes with the applied flux.

\subsection{One-body density matrix and momentum distribution}

Finally, in this section we show that the study of the first-order correlations and momentum distribution allows to obtain information about fragmentation.

\begin{figure}[htbp]
	\centering
	\subfigimg[width=0.24\textwidth]{a}{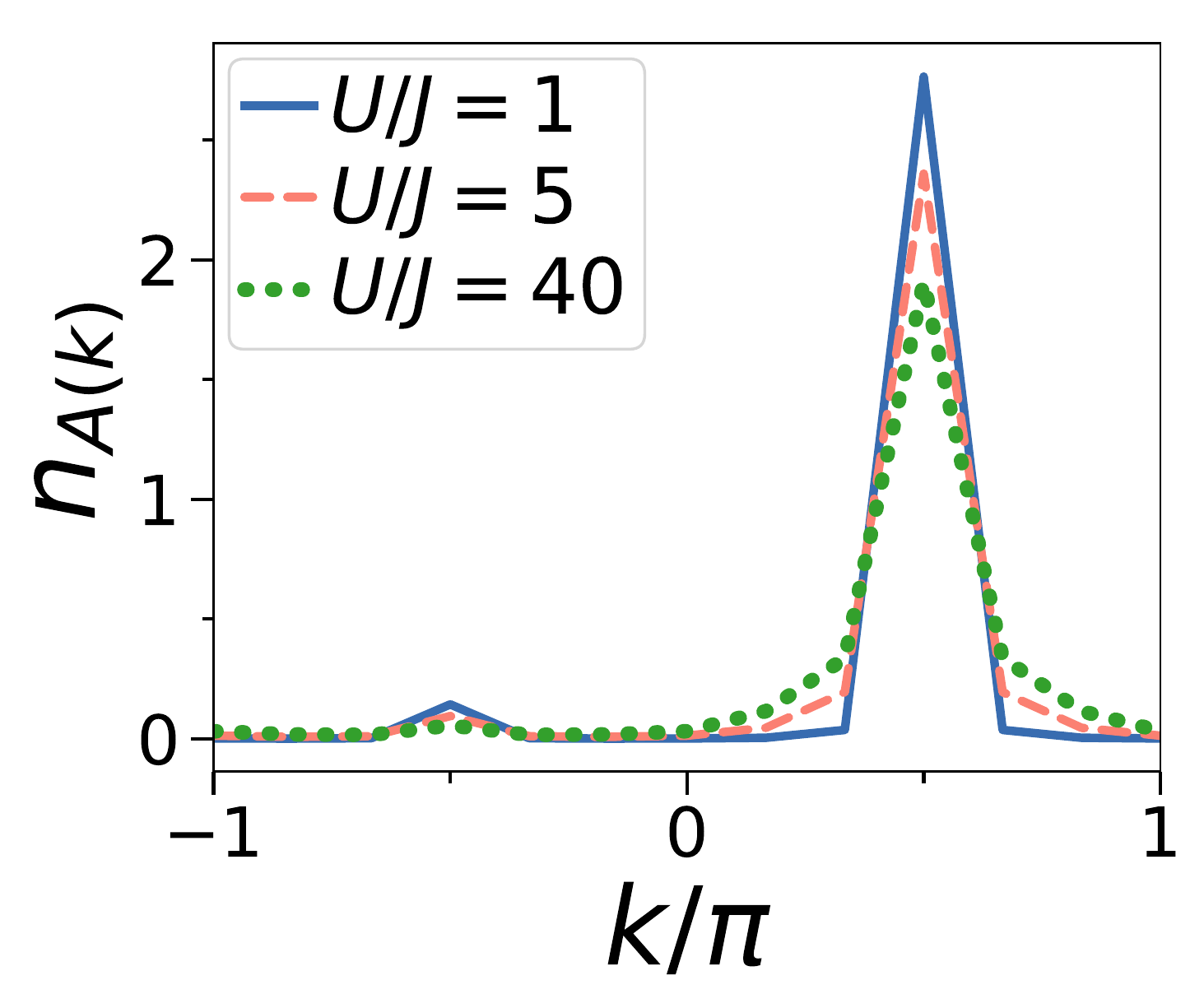}\hfill
	\subfigimg[width=0.24\textwidth]{b}{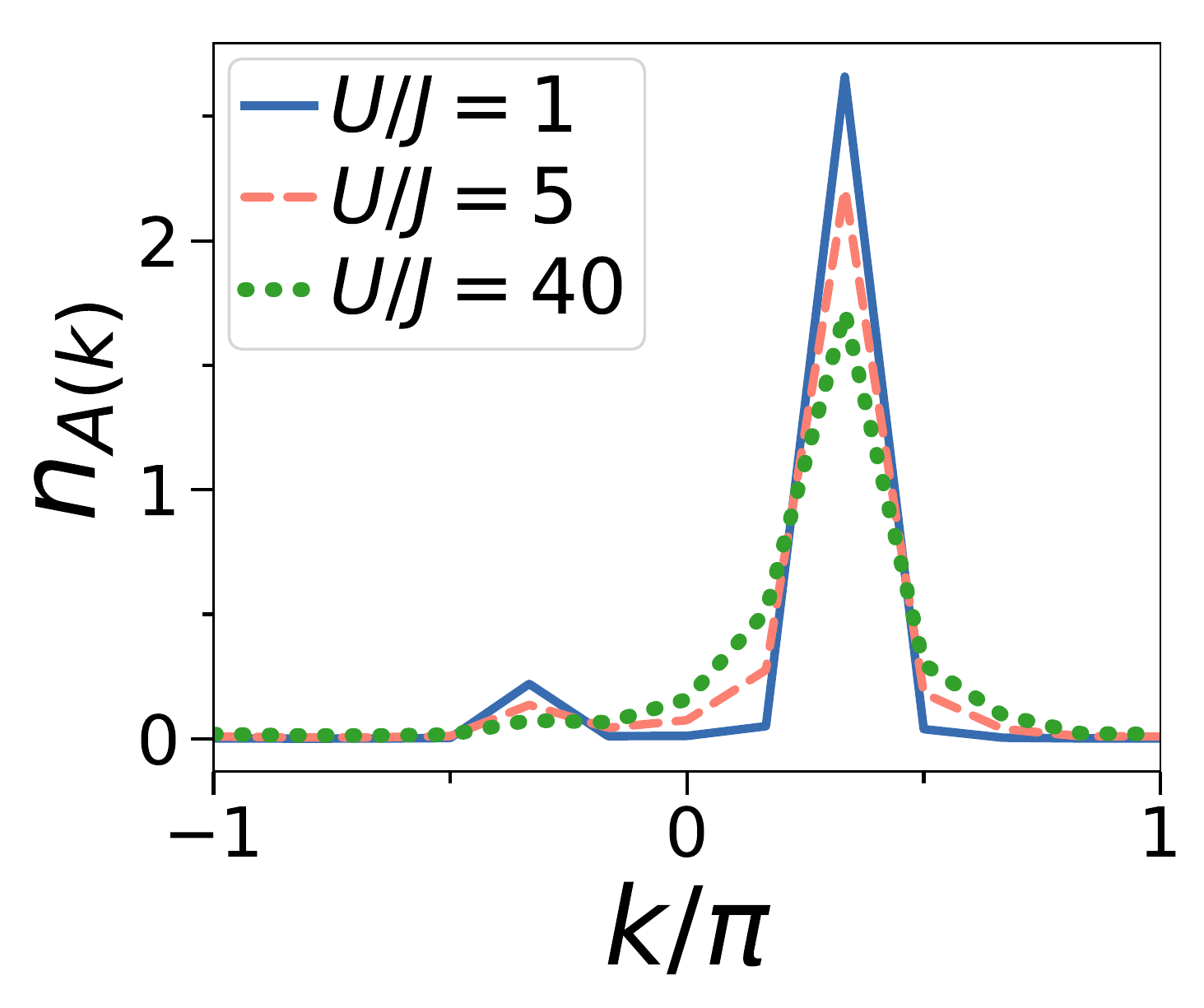}
	
	\caption{Momentum distribution in the leg basis for leg A with \idg{a} $K/J=1$, $\phi=1/4$ \idg{b} $\phi=1/6$. The momentum distribution in ring B is the same graph reflected at zero momentum. The calculation are performed with $N=6$ particles in total in the two rings and ${L=12}$ sites per ring.}
	\label{momentum12}
\end{figure}

The momentum distribution in each ring A and B  is plotted in Fig.\ref{momentum12}. We see that at weak inter-ring coupling the momentum distribution is centered in each of the two minima $k=k_1$ or $k_2$, corresponding to the applied gauge flux on each ring.

\begin{figure}[htbp]
	\centering
	\subfigimg[width=0.23\textwidth]{a}{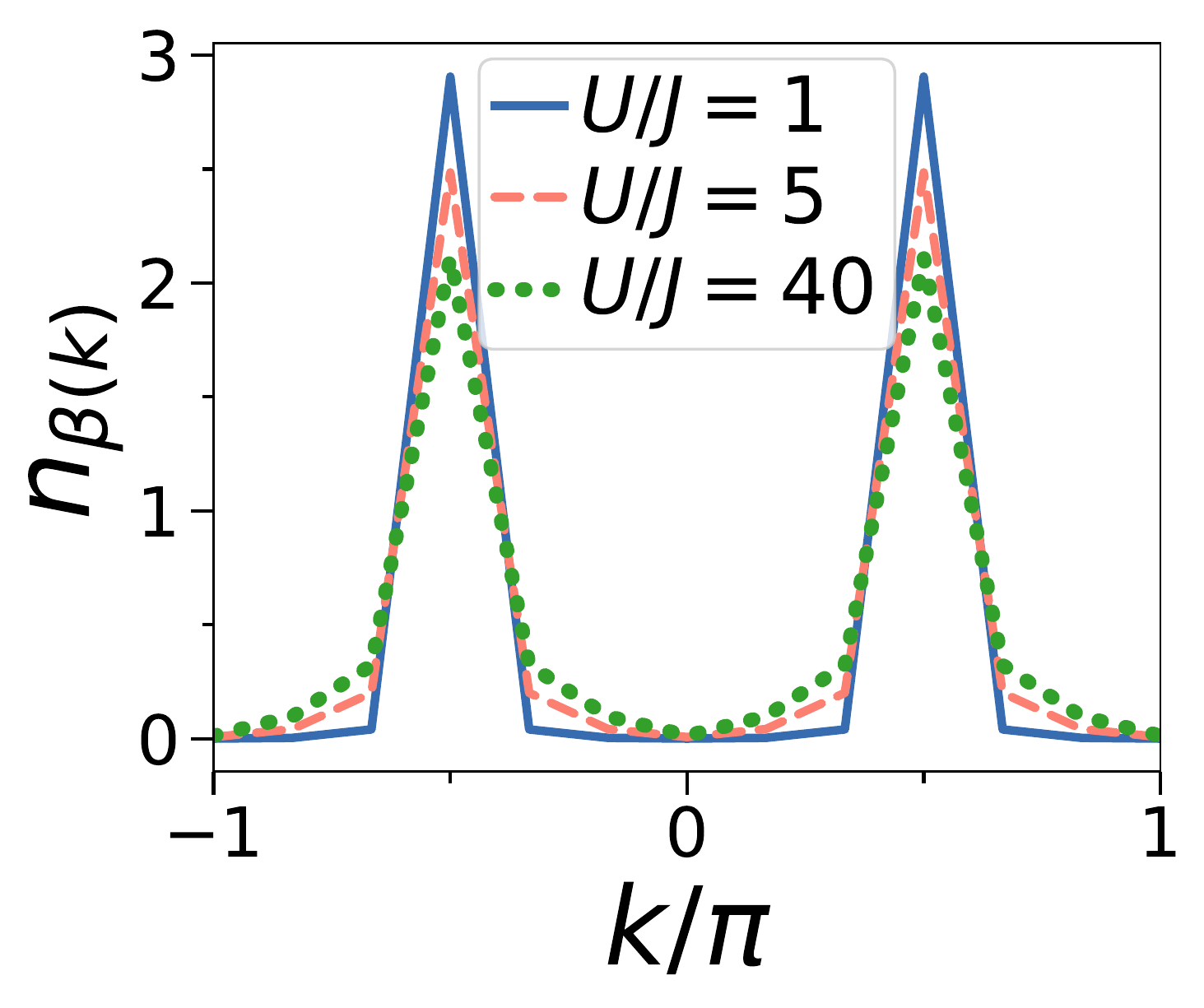}\hfill
	\subfigimg[width=0.25\textwidth]{b}{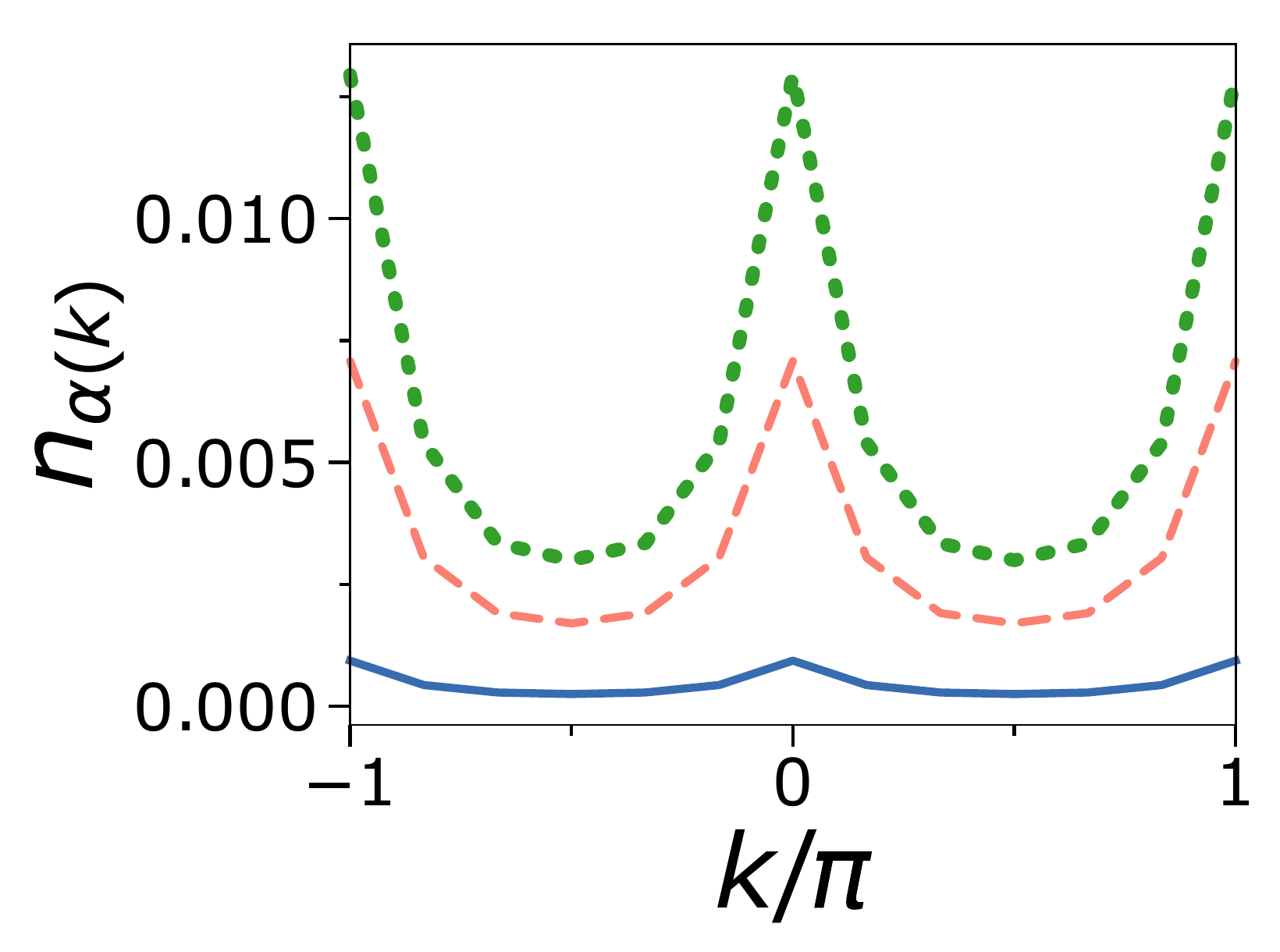}\\
	\subfigimg[width=0.23\textwidth]{c}{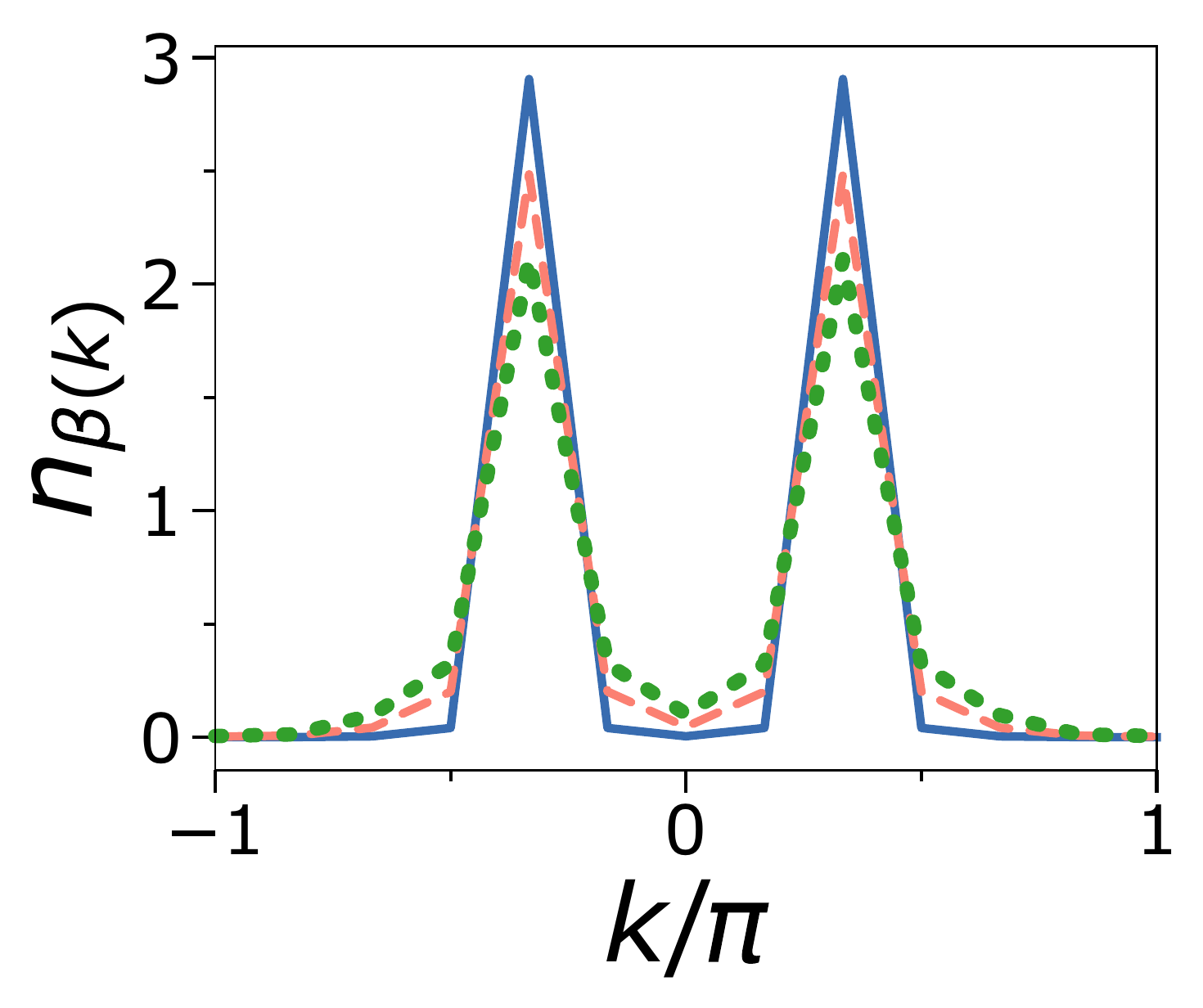}\hfill
	\subfigimg[width=0.25\textwidth]{d}{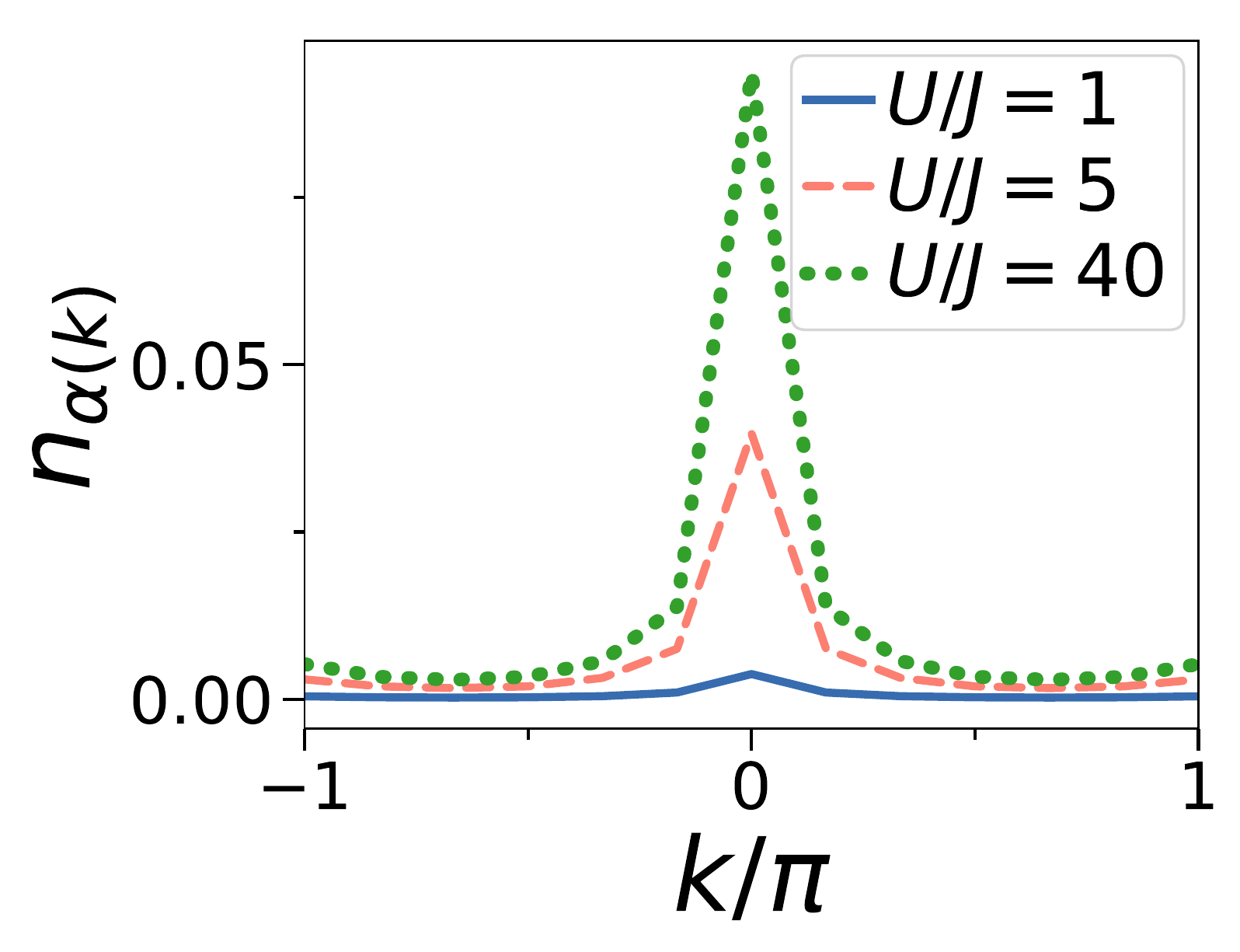}\\
	\caption{Momentum distribution in the diagonal basis for weak inter-ring coupling $K/J=0.1$ with \idg{a,b} $\phi=1/4$ \idg{c,d} $\phi=1/6$. \idg{a,c} shows the lowest branch, \idg{b,d} shows the upper branch. We take $N=6$ particles in total in the two rings and ${L=12}$ sites per ring. }
	\label{momentumSymK01}
\end{figure}

\begin{figure}[htbp]
	\centering
	\subfigimg[width=0.23\textwidth]{a}{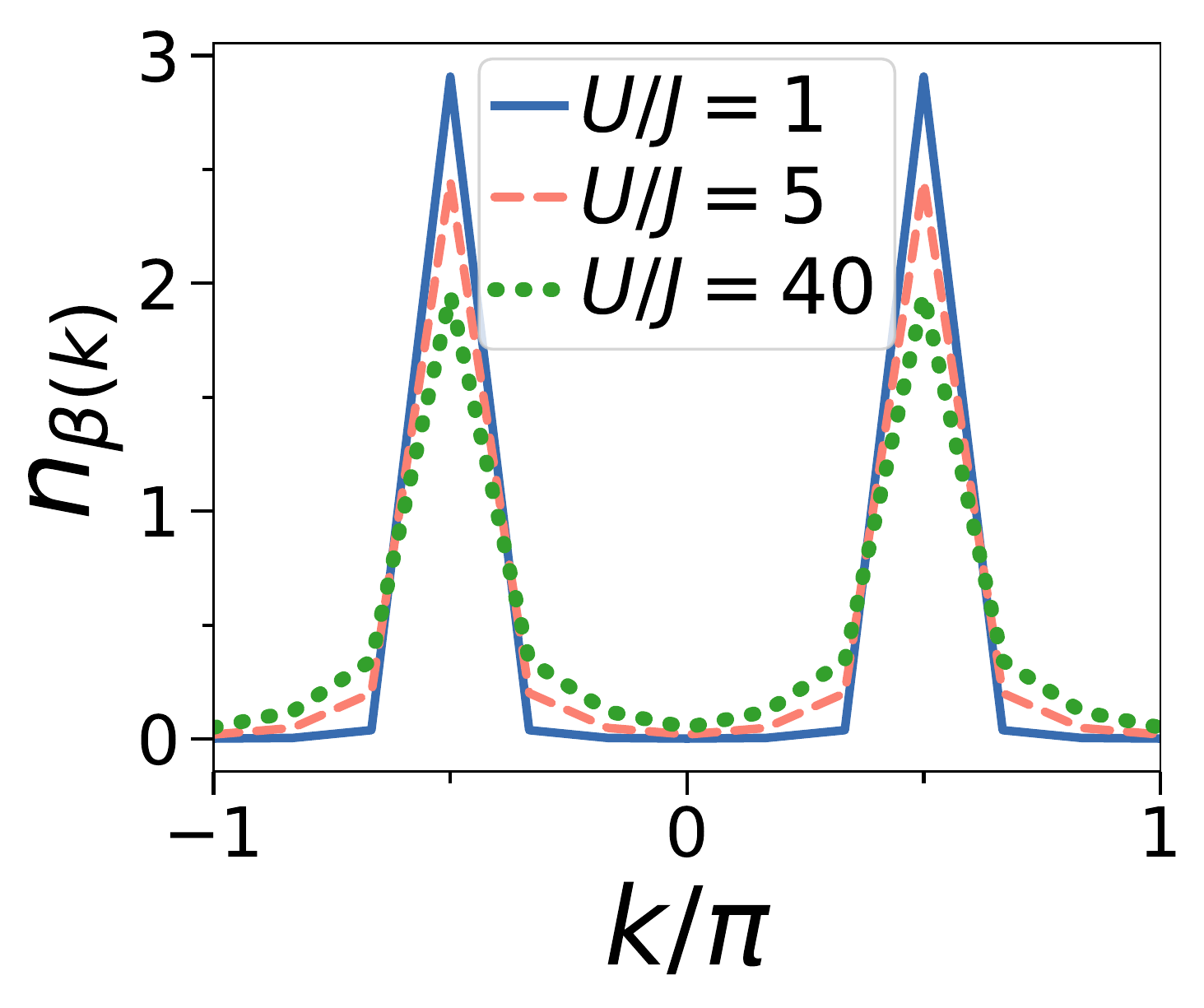}\hfill
	\subfigimg[width=0.25\textwidth]{b}{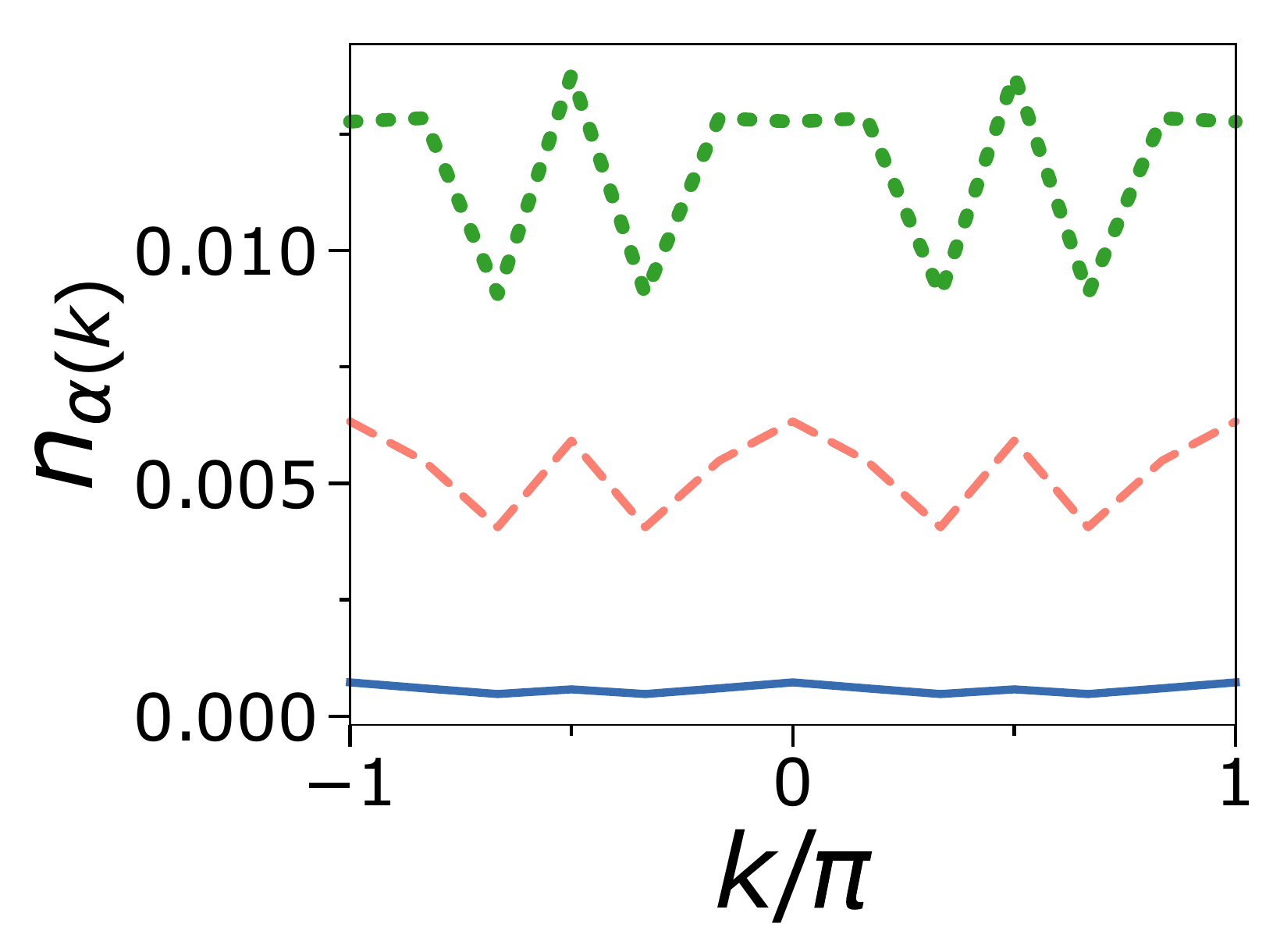}\\
	\subfigimg[width=0.23\textwidth]{c}{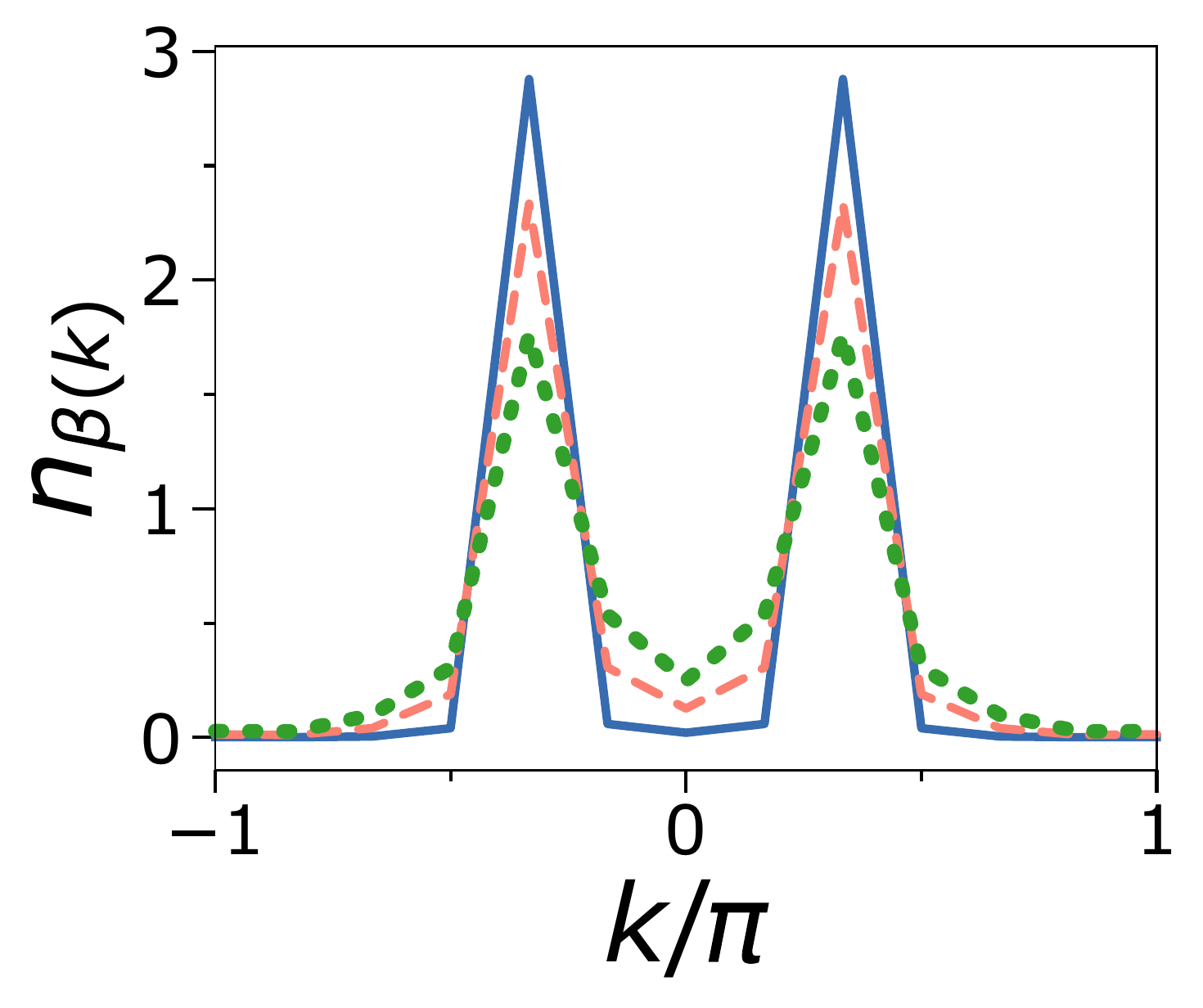}\hfill
	\subfigimg[width=0.25\textwidth]{d}{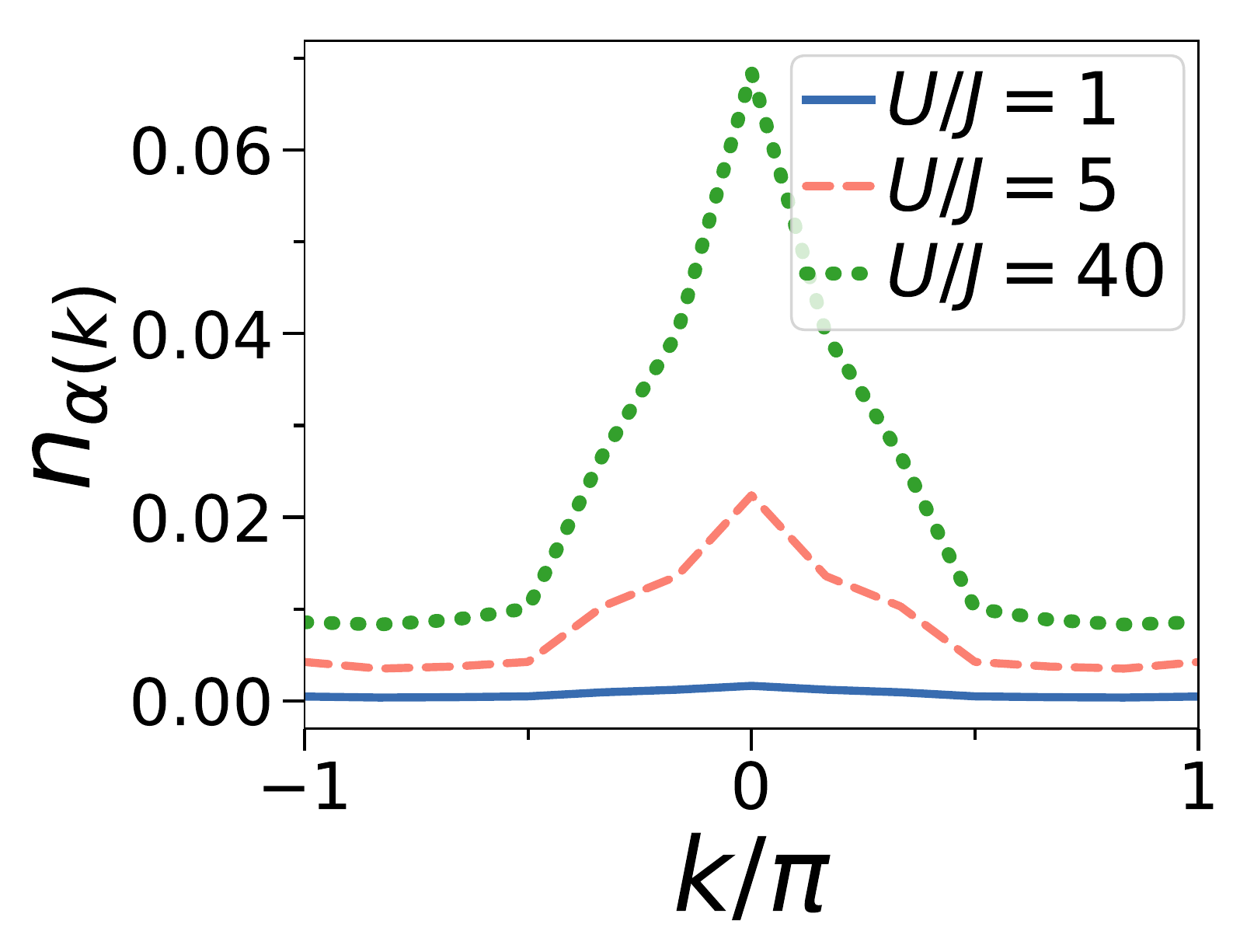}\\
	\caption{Momentum distribution in the diagonal basis for strong inter-ring coupling $K/J=1$ with \idg{a,b} $\phi=1/4$ \idg{c,d} $\phi=1/6$. \idg{a,c} shows the lowest branch, \idg{b,d} shows the upper branch. We take $N=6$ particles in total in the two rings and ${L=12}$ sites per ring. }
	\label{momentumSymK1}
\end{figure}

The momentum distribution in the diagonal basis, corresponding to occupation of lower and upper excitation branch in momentum space,  is plotted in Fig.\ref{momentumSymK01} and \ref{momentumSymK1} for two choices of the inter-ring coupling. It displays a two-peak structure, centered in corresponding of the two minima of the single-particle dispersion relation $k=k_1,k_2$.
We notice first that most of the population occupies the lower branch, while the upper branch population is two orders of magnitude smaller. This validates the reduction to an effective one-dimensional system corresponding to the lower branch discussed in Sec.\ref{fermionization}. Focusing on the lower-branch momentum distribution, we notice that at strong interactions the momentum distribution broadens due to interaction effects as well as develops  large-momentum tails characteristic of strongly-interacting regime. The peaks remain well defined even in the fermionized limit, as typical of bosonic statistics, though their width coincides with the kinetic energy of the corresponding Fermi gas. The width of the momentum distribution for bosons and fermions is shown in the Appendix.

\begin{figure}[htbp]
	\centering
	\subfigimg[width=0.35\textwidth]{}{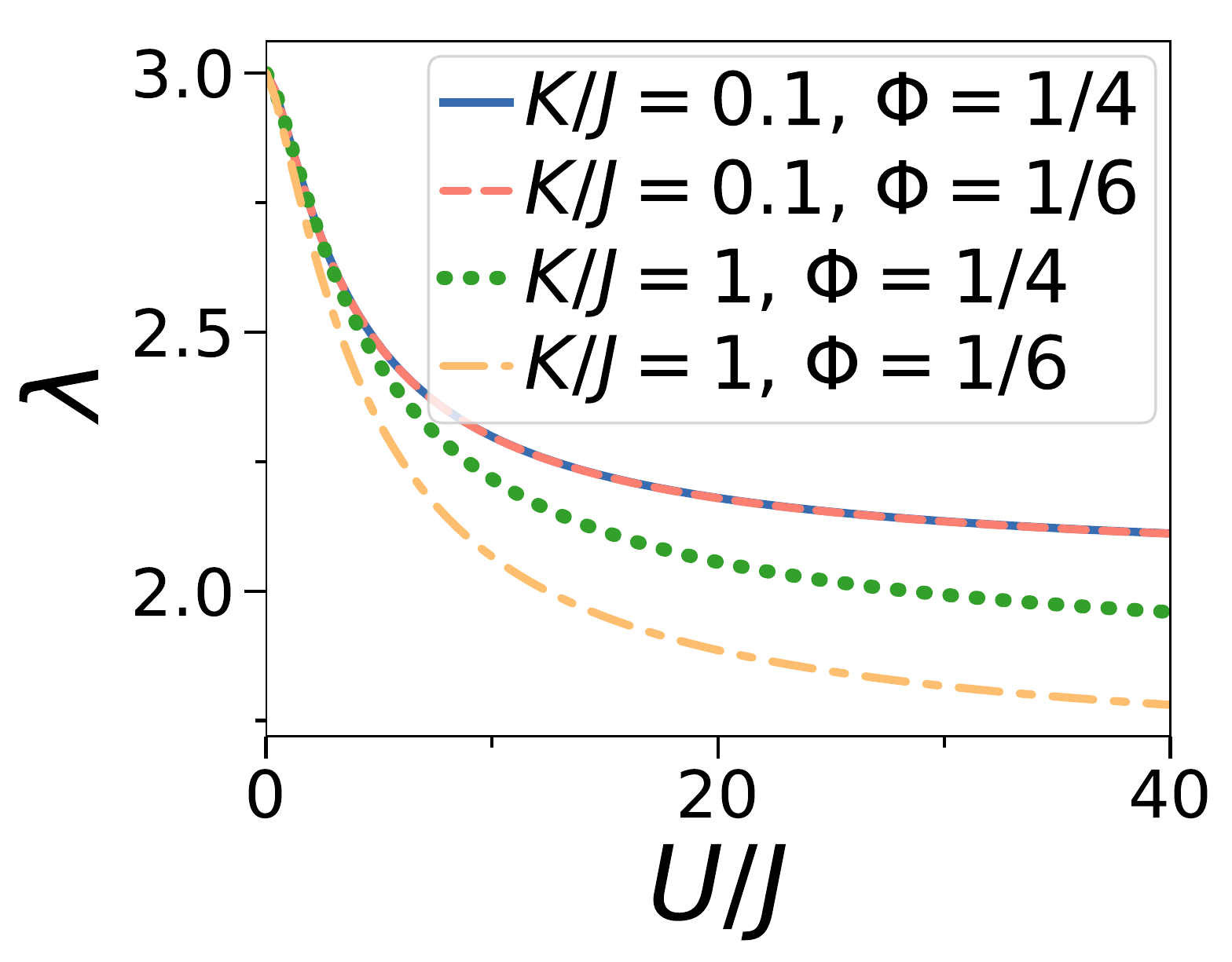}
	\caption{(Color online) Two largest degenerate eigenvalues $\lambda$ of the single particle density matrix as a function of interaction strength $U/J$ for different values of flux $\Phi$ and inter-ring coupling $K/J$ as indicated in the legend. The parameters use in the calculation are  $L=12$, $N=6$.}
	\label{SPDM}
\end{figure}

We finally analyze the single particle density matrix (SPDM), which is independent of the chosen basis and 
whose Fourier transform yields the momentum distribution.  By analyzing its eigenvalues, in the case of weak inter-ring coupling  we find that for all values of interactions, it displays a double degeneracy of the two largest eigenvalues, as shown in Fig.\ref{SPDM}.
The degeneracy of the two largest eigenvalues of the one-body density matrix provides a strong indication of fragmentation.

As already noticed in the study of other observables, for strong inter-ring coupling the quasi-one dimensional description does not apply. We see in particular that the eigenvalues decay faster with increasing interactions, in a flux-dependent way, indicating the important role of the transverse direction in this case.

\section{Summary and concluding remarks}
\label{conclusions}
In this work we have studied the ground-state properties of two tunnel-coupled lattice rings in the quantum regime. In particular, we have shown that  the ground-state of the system is always fragmented at any interaction strength, and the nature of the fragmented state depends on the interaction strength: for weak interactions it consists of fragmentation among two single-particle states, while for strong interactions it corresponds to two fragmented Fermi spheres. 
This description holds provided that the tunnel coupling between the two rings is sufficiently weak and allows for an analytical Ansatz which well describes the limits of very weak or very strong interactions. 
The information of the nature of the state can be inferred by combining the knowledge of various observables: the study of chiral currents and current-current correlation functions allow to identify the vortex phase. By increasing interactions, the flux dependence of the currents across the transitions between states with different winding numbers  is smoothed out - see Fig.\ref{VMNAInt}.  The density-density correlation function shows the onset to  fermionization via the appearance of Friedel-like oscillations at large interactions, and a double-peak structure in the momentum distribution together with the demonstration of degenerate eigenvalues of the one-body density matrix establishes the fragmented nature of the state. 
In outlook, it would be interesting to explore the crossover from quantum regime at very weak filling considered in this work and the  mean-field Gross-Pitaevskii  description used in the case of very large number of bosons per lattice site. 

\acknowledgements
We acknowledge funding from the SuperRing ANR Project (ANR-15-CE30-0012-02). The Grenoble LANEF framework (ANR-10-LABX-51-01) is acknowledged for its support with mutualized infrastructure. We thank National Research Foundation Singapore and the Ministry of Education Singapore Academic Research Fund Tier 2 (Grant No. MOE2015-T2-1-101) for support.

\bibliography{library}

\begin{thebibliography}{51}%
\makeatletter
\providecommand \@ifxundefined [1]{%
 \@ifx{#1\undefined}
}%
\providecommand \@ifnum [1]{%
 \ifnum #1\expandafter \@firstoftwo
 \else \expandafter \@secondoftwo
 \fi
}%
\providecommand \@ifx [1]{%
 \ifx #1\expandafter \@firstoftwo
 \else \expandafter \@secondoftwo
 \fi
}%
\providecommand \natexlab [1]{#1}%
\providecommand \enquote  [1]{``#1''}%
\providecommand \bibnamefont  [1]{#1}%
\providecommand \bibfnamefont [1]{#1}%
\providecommand \citenamefont [1]{#1}%
\providecommand \href@noop [0]{\@secondoftwo}%
\providecommand \href [0]{\begingroup \@sanitize@url \@href}%
\providecommand \@href[1]{\@@startlink{#1}\@@href}%
\providecommand \@@href[1]{\endgroup#1\@@endlink}%
\providecommand \@sanitize@url [0]{\catcode `\\12\catcode `\$12\catcode
  `\&12\catcode `\#12\catcode `\^12\catcode `\_12\catcode `\%12\relax}%
\providecommand \@@startlink[1]{}%
\providecommand \@@endlink[0]{}%
\providecommand \url  [0]{\begingroup\@sanitize@url \@url }%
\providecommand \@url [1]{\endgroup\@href {#1}{\urlprefix }}%
\providecommand \urlprefix  [0]{URL }%
\providecommand \Eprint [0]{\href }%
\providecommand \doibase [0]{http://dx.doi.org/}%
\providecommand \selectlanguage [0]{\@gobble}%
\providecommand \bibinfo  [0]{\@secondoftwo}%
\providecommand \bibfield  [0]{\@secondoftwo}%
\providecommand \translation [1]{[#1]}%
\providecommand \BibitemOpen [0]{}%
\providecommand \bibitemStop [0]{}%
\providecommand \bibitemNoStop [0]{.\EOS\space}%
\providecommand \EOS [0]{\spacefactor3000\relax}%
\providecommand \BibitemShut  [1]{\csname bibitem#1\endcsname}%
\let\auto@bib@innerbib\@empty
\bibitem [{\citenamefont {Bloch}\ \emph {et~al.}(2008)\citenamefont {Bloch},
  \citenamefont {Dalibard},\ and\ \citenamefont {Zwerger}}]{bloch2008many}%
  \BibitemOpen
  \bibfield  {author} {\bibinfo {author} {\bibfnamefont {I.}~\bibnamefont
  {Bloch}}, \bibinfo {author} {\bibfnamefont {J.}~\bibnamefont {Dalibard}}, \
  and\ \bibinfo {author} {\bibfnamefont {W.}~\bibnamefont {Zwerger}},\
  }\href@noop {} {\bibfield  {journal} {\bibinfo  {journal} {Rev. Mod. Phys.}\
  }\textbf {\bibinfo {volume} {80}},\ \bibinfo {pages} {885} (\bibinfo {year}
  {2008})}\BibitemShut {NoStop}%
\bibitem [{\citenamefont {Kardar}(1986)}]{kardar1986josephson}%
  \BibitemOpen
  \bibfield  {author} {\bibinfo {author} {\bibfnamefont {M.}~\bibnamefont
  {Kardar}},\ }\href@noop {} {\bibfield  {journal} {\bibinfo  {journal} {Phys.
  Rev. B}\ }\textbf {\bibinfo {volume} {33}},\ \bibinfo {pages} {3125}
  (\bibinfo {year} {1986})}\BibitemShut {NoStop}%
\bibitem [{\citenamefont {Orignac}\ and\ \citenamefont
  {Giamarchi}(2001)}]{orignac2001meissner}%
  \BibitemOpen
  \bibfield  {author} {\bibinfo {author} {\bibfnamefont {E.}~\bibnamefont
  {Orignac}}\ and\ \bibinfo {author} {\bibfnamefont {T.}~\bibnamefont
  {Giamarchi}},\ }\href@noop {} {\bibfield  {journal} {\bibinfo  {journal}
  {Phys. Rev. B}\ }\textbf {\bibinfo {volume} {64}},\ \bibinfo {pages} {144515}
  (\bibinfo {year} {2001})}\BibitemShut {NoStop}%
\bibitem [{\citenamefont {Petrescu}\ and\ \citenamefont
  {Le~Hur}(2015)}]{petrescu2015chiral}%
  \BibitemOpen
  \bibfield  {author} {\bibinfo {author} {\bibfnamefont {A.}~\bibnamefont
  {Petrescu}}\ and\ \bibinfo {author} {\bibfnamefont {K.}~\bibnamefont
  {Le~Hur}},\ }\href@noop {} {\bibfield  {journal} {\bibinfo  {journal}
  {Physical Review B}\ }\textbf {\bibinfo {volume} {91}},\ \bibinfo {pages}
  {054520} (\bibinfo {year} {2015})}\BibitemShut {NoStop}%
\bibitem [{\citenamefont {Atala}\ \emph {et~al.}(2014)\citenamefont {Atala},
  \citenamefont {Aidelsburger}, \citenamefont {Lohse}, \citenamefont
  {Barreiro}, \citenamefont {Paredes},\ and\ \citenamefont
  {Bloch}}]{atala2014observation}%
  \BibitemOpen
  \bibfield  {author} {\bibinfo {author} {\bibfnamefont {M.}~\bibnamefont
  {Atala}}, \bibinfo {author} {\bibfnamefont {M.}~\bibnamefont {Aidelsburger}},
  \bibinfo {author} {\bibfnamefont {M.}~\bibnamefont {Lohse}}, \bibinfo
  {author} {\bibfnamefont {J.~T.}\ \bibnamefont {Barreiro}}, \bibinfo {author}
  {\bibfnamefont {B.}~\bibnamefont {Paredes}}, \ and\ \bibinfo {author}
  {\bibfnamefont {I.}~\bibnamefont {Bloch}},\ }\href@noop {} {\bibfield
  {journal} {\bibinfo  {journal} {Nat. Phys.}\ }\textbf {\bibinfo {volume}
  {10}},\ \bibinfo {pages} {588} (\bibinfo {year} {2014})}\BibitemShut
  {NoStop}%
\bibitem [{\citenamefont {Mancini}\ \emph {et~al.}(2015)\citenamefont
  {Mancini}, \citenamefont {Pagano}, \citenamefont {Cappellini}, \citenamefont
  {Livi}, \citenamefont {Rider}, \citenamefont {Catani}, \citenamefont {Sias},
  \citenamefont {Zoller}, \citenamefont {Inguscio}, \citenamefont {Dalmonte}
  \emph {et~al.}}]{mancini2015observation}%
  \BibitemOpen
  \bibfield  {author} {\bibinfo {author} {\bibfnamefont {M.}~\bibnamefont
  {Mancini}}, \bibinfo {author} {\bibfnamefont {G.}~\bibnamefont {Pagano}},
  \bibinfo {author} {\bibfnamefont {G.}~\bibnamefont {Cappellini}}, \bibinfo
  {author} {\bibfnamefont {L.}~\bibnamefont {Livi}}, \bibinfo {author}
  {\bibfnamefont {M.}~\bibnamefont {Rider}}, \bibinfo {author} {\bibfnamefont
  {J.}~\bibnamefont {Catani}}, \bibinfo {author} {\bibfnamefont
  {C.}~\bibnamefont {Sias}}, \bibinfo {author} {\bibfnamefont {P.}~\bibnamefont
  {Zoller}}, \bibinfo {author} {\bibfnamefont {M.}~\bibnamefont {Inguscio}},
  \bibinfo {author} {\bibfnamefont {M.}~\bibnamefont {Dalmonte}},  \emph
  {et~al.},\ }\href@noop {} {\bibfield  {journal} {\bibinfo  {journal}
  {Science}\ }\textbf {\bibinfo {volume} {349}},\ \bibinfo {pages} {1510}
  (\bibinfo {year} {2015})}\BibitemShut {NoStop}%
\bibitem [{\citenamefont {Stuhl}\ \emph {et~al.}(2015)\citenamefont {Stuhl},
  \citenamefont {Lu}, \citenamefont {Aycock}, \citenamefont {Genkina},\ and\
  \citenamefont {Spielman}}]{stuhl2015visualizing}%
  \BibitemOpen
  \bibfield  {author} {\bibinfo {author} {\bibfnamefont {B.}~\bibnamefont
  {Stuhl}}, \bibinfo {author} {\bibfnamefont {H.-I.}\ \bibnamefont {Lu}},
  \bibinfo {author} {\bibfnamefont {L.}~\bibnamefont {Aycock}}, \bibinfo
  {author} {\bibfnamefont {D.}~\bibnamefont {Genkina}}, \ and\ \bibinfo
  {author} {\bibfnamefont {I.}~\bibnamefont {Spielman}},\ }\href@noop {}
  {\bibfield  {journal} {\bibinfo  {journal} {Science}\ }\textbf {\bibinfo
  {volume} {349}},\ \bibinfo {pages} {1514} (\bibinfo {year}
  {2015})}\BibitemShut {NoStop}%
\bibitem [{\citenamefont {An}\ \emph {et~al.}()\citenamefont {An},
  \citenamefont {Meier},\ and\ \citenamefont {Gadway}}]{an2016direct}%
  \BibitemOpen
  \bibfield  {author} {\bibinfo {author} {\bibfnamefont {F.~A.}\ \bibnamefont
  {An}}, \bibinfo {author} {\bibfnamefont {E.~J.}\ \bibnamefont {Meier}}, \
  and\ \bibinfo {author} {\bibfnamefont {B.}~\bibnamefont {Gadway}},\
  }\href@noop {} {\bibinfo  {journal} {arXiv:1609.09467}\ }\BibitemShut
  {NoStop}%
\bibitem [{\citenamefont {Livi}\ \emph {et~al.}(2016)\citenamefont {Livi},
  \citenamefont {Cappellini}, \citenamefont {Diem}, \citenamefont {Franchi},
  \citenamefont {Clivati}, \citenamefont {Frittelli}, \citenamefont {Levi},
  \citenamefont {Calonico}, \citenamefont {Catani}, \citenamefont {Inguscio}
  \emph {et~al.}}]{livi2016synthetic}%
  \BibitemOpen
\bibfield  {journal} {  }\bibfield  {author} {\bibinfo {author} {\bibfnamefont
  {L.}~\bibnamefont {Livi}}, \bibinfo {author} {\bibfnamefont {G.}~\bibnamefont
  {Cappellini}}, \bibinfo {author} {\bibfnamefont {M.}~\bibnamefont {Diem}},
  \bibinfo {author} {\bibfnamefont {L.}~\bibnamefont {Franchi}}, \bibinfo
  {author} {\bibfnamefont {C.}~\bibnamefont {Clivati}}, \bibinfo {author}
  {\bibfnamefont {M.}~\bibnamefont {Frittelli}}, \bibinfo {author}
  {\bibfnamefont {F.}~\bibnamefont {Levi}}, \bibinfo {author} {\bibfnamefont
  {D.}~\bibnamefont {Calonico}}, \bibinfo {author} {\bibfnamefont
  {J.}~\bibnamefont {Catani}}, \bibinfo {author} {\bibfnamefont
  {M.}~\bibnamefont {Inguscio}},  \emph {et~al.},\ }\href@noop {} {\bibfield
  {journal} {\bibinfo  {journal} {Phys. Rev. Lett.}\ }\textbf {\bibinfo
  {volume} {117}},\ \bibinfo {pages} {220401} (\bibinfo {year}
  {2016})}\BibitemShut {NoStop}%
\bibitem [{\citenamefont {Petrescu}\ and\ \citenamefont
  {Le~Hur}(2013)}]{petrescu2013bosonic}%
  \BibitemOpen
  \bibfield  {author} {\bibinfo {author} {\bibfnamefont {A.}~\bibnamefont
  {Petrescu}}\ and\ \bibinfo {author} {\bibfnamefont {K.}~\bibnamefont
  {Le~Hur}},\ }\href@noop {} {\bibfield  {journal} {\bibinfo  {journal} {Phys.
  Rev. Lett.}\ }\textbf {\bibinfo {volume} {111}},\ \bibinfo {pages} {150601}
  (\bibinfo {year} {2013})}\BibitemShut {NoStop}%
\bibitem [{\citenamefont {Tokuno}\ and\ \citenamefont
  {Georges}(2014)}]{tokuno2014ground}%
  \BibitemOpen
  \bibfield  {author} {\bibinfo {author} {\bibfnamefont {A.}~\bibnamefont
  {Tokuno}}\ and\ \bibinfo {author} {\bibfnamefont {A.}~\bibnamefont
  {Georges}},\ }\href@noop {} {\bibfield  {journal} {\bibinfo  {journal} {New
  J. Phys.}\ }\textbf {\bibinfo {volume} {16}},\ \bibinfo {pages} {073005}
  (\bibinfo {year} {2014})}\BibitemShut {NoStop}%
\bibitem [{\citenamefont {Wei}\ and\ \citenamefont
  {Mueller}(2014)}]{wei2014theory}%
  \BibitemOpen
  \bibfield  {author} {\bibinfo {author} {\bibfnamefont {R.}~\bibnamefont
  {Wei}}\ and\ \bibinfo {author} {\bibfnamefont {E.~J.}\ \bibnamefont
  {Mueller}},\ }\href@noop {} {\bibfield  {journal} {\bibinfo  {journal} {Phys.
  Rev. A}\ }\textbf {\bibinfo {volume} {89}},\ \bibinfo {pages} {063617}
  (\bibinfo {year} {2014})}\BibitemShut {NoStop}%
\bibitem [{\citenamefont {Di~Dio}\ \emph {et~al.}(2015)\citenamefont {Di~Dio},
  \citenamefont {Citro}, \citenamefont {De~Palo}, \citenamefont {Orignac},\
  and\ \citenamefont {Chiofalo}}]{di2015meissner}%
  \BibitemOpen
  \bibfield  {author} {\bibinfo {author} {\bibfnamefont {M.}~\bibnamefont
  {Di~Dio}}, \bibinfo {author} {\bibfnamefont {R.}~\bibnamefont {Citro}},
  \bibinfo {author} {\bibfnamefont {S.}~\bibnamefont {De~Palo}}, \bibinfo
  {author} {\bibfnamefont {E.}~\bibnamefont {Orignac}}, \ and\ \bibinfo
  {author} {\bibfnamefont {M.-L.}\ \bibnamefont {Chiofalo}},\ }\href@noop {}
  {\bibfield  {journal} {\bibinfo  {journal} {Eur. Phys. J. Special Topics}\
  }\textbf {\bibinfo {volume} {224}},\ \bibinfo {pages} {525} (\bibinfo {year}
  {2015})}\BibitemShut {NoStop}%
\bibitem [{\citenamefont {Orignac}\ \emph {et~al.}(2016)\citenamefont
  {Orignac}, \citenamefont {Citro}, \citenamefont {Di~Dio}, \citenamefont
  {De~Palo},\ and\ \citenamefont {Chiofalo}}]{orignac2016incommensurate}%
  \BibitemOpen
  \bibfield  {author} {\bibinfo {author} {\bibfnamefont {E.}~\bibnamefont
  {Orignac}}, \bibinfo {author} {\bibfnamefont {R.}~\bibnamefont {Citro}},
  \bibinfo {author} {\bibfnamefont {M.}~\bibnamefont {Di~Dio}}, \bibinfo
  {author} {\bibfnamefont {S.}~\bibnamefont {De~Palo}}, \ and\ \bibinfo
  {author} {\bibfnamefont {M.-L.}\ \bibnamefont {Chiofalo}},\ }\href@noop {}
  {\bibfield  {journal} {\bibinfo  {journal} {New J. Phys.}\ }\textbf {\bibinfo
  {volume} {18}},\ \bibinfo {pages} {055017} (\bibinfo {year}
  {2016})}\BibitemShut {NoStop}%
\bibitem [{\citenamefont {Greschner}\ \emph {et~al.}(2015)\citenamefont
  {Greschner}, \citenamefont {Piraud}, \citenamefont {Heidrich-Meisner},
  \citenamefont {McCulloch}, \citenamefont {Schollw{\"o}ck},\ and\
  \citenamefont {Vekua}}]{greschner2015spontaneous}%
  \BibitemOpen
  \bibfield  {author} {\bibinfo {author} {\bibfnamefont {S.}~\bibnamefont
  {Greschner}}, \bibinfo {author} {\bibfnamefont {M.}~\bibnamefont {Piraud}},
  \bibinfo {author} {\bibfnamefont {F.}~\bibnamefont {Heidrich-Meisner}},
  \bibinfo {author} {\bibfnamefont {I.}~\bibnamefont {McCulloch}}, \bibinfo
  {author} {\bibfnamefont {U.}~\bibnamefont {Schollw{\"o}ck}}, \ and\ \bibinfo
  {author} {\bibfnamefont {T.}~\bibnamefont {Vekua}},\ }\href@noop {}
  {\bibfield  {journal} {\bibinfo  {journal} {Phys. Rev. Lett.}\ }\textbf
  {\bibinfo {volume} {115}},\ \bibinfo {pages} {190402} (\bibinfo {year}
  {2015})}\BibitemShut {NoStop}%
\bibitem [{\citenamefont {Piraud}\ \emph {et~al.}(2015)\citenamefont {Piraud},
  \citenamefont {Heidrich-Meisner}, \citenamefont {McCulloch}, \citenamefont
  {Greschner}, \citenamefont {Vekua},\ and\ \citenamefont
  {Schollw{\"o}ck}}]{piraud2015vortex}%
  \BibitemOpen
  \bibfield  {author} {\bibinfo {author} {\bibfnamefont {M.}~\bibnamefont
  {Piraud}}, \bibinfo {author} {\bibfnamefont {F.}~\bibnamefont
  {Heidrich-Meisner}}, \bibinfo {author} {\bibfnamefont {I.~P.}\ \bibnamefont
  {McCulloch}}, \bibinfo {author} {\bibfnamefont {S.}~\bibnamefont
  {Greschner}}, \bibinfo {author} {\bibfnamefont {T.}~\bibnamefont {Vekua}}, \
  and\ \bibinfo {author} {\bibfnamefont {U.}~\bibnamefont {Schollw{\"o}ck}},\
  }\href@noop {} {\bibfield  {journal} {\bibinfo  {journal} {Phys. Rev. B}\
  }\textbf {\bibinfo {volume} {91}},\ \bibinfo {pages} {140406} (\bibinfo
  {year} {2015})}\BibitemShut {NoStop}%
\bibitem [{\citenamefont {Dalibard}\ \emph {et~al.}(2011)\citenamefont
  {Dalibard}, \citenamefont {Gerbier}, \citenamefont {Juzeli{\=u}nas},\ and\
  \citenamefont {{\"O}hberg}}]{dalibard2011colloquium}%
  \BibitemOpen
  \bibfield  {author} {\bibinfo {author} {\bibfnamefont {J.}~\bibnamefont
  {Dalibard}}, \bibinfo {author} {\bibfnamefont {F.}~\bibnamefont {Gerbier}},
  \bibinfo {author} {\bibfnamefont {G.}~\bibnamefont {Juzeli{\=u}nas}}, \ and\
  \bibinfo {author} {\bibfnamefont {P.}~\bibnamefont {{\"O}hberg}},\
  }\href@noop {} {\bibfield  {journal} {\bibinfo  {journal} {Rev. Mod. Phys.}\
  }\textbf {\bibinfo {volume} {83}},\ \bibinfo {pages} {1523} (\bibinfo {year}
  {2011})}\BibitemShut {NoStop}%
\bibitem [{\citenamefont {Wright}\ \emph {et~al.}(2013)\citenamefont {Wright},
  \citenamefont {Blakestad}, \citenamefont {Lobb}, \citenamefont {Phillips},\
  and\ \citenamefont {Campbell}}]{wright2013driving}%
  \BibitemOpen
  \bibfield  {author} {\bibinfo {author} {\bibfnamefont {K.~C.}\ \bibnamefont
  {Wright}}, \bibinfo {author} {\bibfnamefont {R.~B.}\ \bibnamefont
  {Blakestad}}, \bibinfo {author} {\bibfnamefont {C.~J.}\ \bibnamefont {Lobb}},
  \bibinfo {author} {\bibfnamefont {W.~D.}\ \bibnamefont {Phillips}}, \ and\
  \bibinfo {author} {\bibfnamefont {G.~K.}\ \bibnamefont {Campbell}},\
  }\href@noop {} {\bibfield  {journal} {\bibinfo  {journal} {Phys. Rev. Lett.}\
  }\textbf {\bibinfo {volume} {110}},\ \bibinfo {pages} {025302} (\bibinfo
  {year} {2013})}\BibitemShut {NoStop}%
\bibitem [{\citenamefont {Ramanathan}\ \emph {et~al.}(2011)\citenamefont
  {Ramanathan}, \citenamefont {Wright}, \citenamefont {Muniz}, \citenamefont
  {Zelan}, \citenamefont {Hill}, \citenamefont {Lobb}, \citenamefont
  {Helmerson}, \citenamefont {Phillips},\ and\ \citenamefont
  {Campbell}}]{Ramanathan2011}%
  \BibitemOpen
  \bibfield  {author} {\bibinfo {author} {\bibfnamefont {A.}~\bibnamefont
  {Ramanathan}}, \bibinfo {author} {\bibfnamefont {K.~C.}\ \bibnamefont
  {Wright}}, \bibinfo {author} {\bibfnamefont {S.~R.}\ \bibnamefont {Muniz}},
  \bibinfo {author} {\bibfnamefont {M.}~\bibnamefont {Zelan}}, \bibinfo
  {author} {\bibfnamefont {W.~T.}\ \bibnamefont {Hill}}, \bibinfo {author}
  {\bibfnamefont {C.~J.}\ \bibnamefont {Lobb}}, \bibinfo {author}
  {\bibfnamefont {K.}~\bibnamefont {Helmerson}}, \bibinfo {author}
  {\bibfnamefont {W.~D.}\ \bibnamefont {Phillips}}, \ and\ \bibinfo {author}
  {\bibfnamefont {G.~K.}\ \bibnamefont {Campbell}},\ }\href {\doibase
  10.1103/PhysRevLett.106.130401} {\bibfield  {journal} {\bibinfo  {journal}
  {Phys. Rev. Lett.}\ }\textbf {\bibinfo {volume} {106}},\ \bibinfo {pages}
  {130401} (\bibinfo {year} {2011})}\BibitemShut {NoStop}%
\bibitem [{\citenamefont {Ryu}\ \emph {et~al.}(2013)\citenamefont {Ryu},
  \citenamefont {Blackburn}, \citenamefont {Blinova},\ and\ \citenamefont
  {Boshier}}]{Ryu2013}%
  \BibitemOpen
  \bibfield  {author} {\bibinfo {author} {\bibfnamefont {C.}~\bibnamefont
  {Ryu}}, \bibinfo {author} {\bibfnamefont {P.~W.}\ \bibnamefont {Blackburn}},
  \bibinfo {author} {\bibfnamefont {A.~A.}\ \bibnamefont {Blinova}}, \ and\
  \bibinfo {author} {\bibfnamefont {M.~G.}\ \bibnamefont {Boshier}},\ }\href
  {\doibase 10.1103/PhysRevLett.111.205301} {\bibfield  {journal} {\bibinfo
  {journal} {Phys. Rev. Lett.}\ }\textbf {\bibinfo {volume} {111}},\ \bibinfo
  {pages} {205301} (\bibinfo {year} {2013})}\BibitemShut {NoStop}%
\bibitem [{\citenamefont {Eckel}\ \emph {et~al.}(2014)\citenamefont {Eckel},
  \citenamefont {Lee}, \citenamefont {Jendrzejewski}, \citenamefont {Murray},
  \citenamefont {Clark}, \citenamefont {Lobb}, \citenamefont {Phillips},
  \citenamefont {Edwards},\ and\ \citenamefont
  {Campbell}}]{eckel2014hysteresis}%
  \BibitemOpen
  \bibfield  {author} {\bibinfo {author} {\bibfnamefont {S.}~\bibnamefont
  {Eckel}}, \bibinfo {author} {\bibfnamefont {J.~G.}\ \bibnamefont {Lee}},
  \bibinfo {author} {\bibfnamefont {F.}~\bibnamefont {Jendrzejewski}}, \bibinfo
  {author} {\bibfnamefont {N.}~\bibnamefont {Murray}}, \bibinfo {author}
  {\bibfnamefont {C.~W.}\ \bibnamefont {Clark}}, \bibinfo {author}
  {\bibfnamefont {C.~J.}\ \bibnamefont {Lobb}}, \bibinfo {author}
  {\bibfnamefont {W.~D.}\ \bibnamefont {Phillips}}, \bibinfo {author}
  {\bibfnamefont {M.}~\bibnamefont {Edwards}}, \ and\ \bibinfo {author}
  {\bibfnamefont {G.~K.}\ \bibnamefont {Campbell}},\ }\href@noop {} {\bibfield
  {journal} {\bibinfo  {journal} {Nature}\ }\textbf {\bibinfo {volume} {506}},\
  \bibinfo {pages} {200} (\bibinfo {year} {2014})}\BibitemShut {NoStop}%
\bibitem [{\citenamefont {Yakimenko}\ \emph {et~al.}(2015)\citenamefont
  {Yakimenko}, \citenamefont {Bidasyuk}, \citenamefont {Weyrauch},
  \citenamefont {Kuriatnikov},\ and\ \citenamefont
  {Vilchinskii}}]{yakimenko2015}%
  \BibitemOpen
  \bibfield  {author} {\bibinfo {author} {\bibfnamefont {A.~I.}\ \bibnamefont
  {Yakimenko}}, \bibinfo {author} {\bibfnamefont {Y.~M.}\ \bibnamefont
  {Bidasyuk}}, \bibinfo {author} {\bibfnamefont {M.}~\bibnamefont {Weyrauch}},
  \bibinfo {author} {\bibfnamefont {Y.~I.}\ \bibnamefont {Kuriatnikov}}, \ and\
  \bibinfo {author} {\bibfnamefont {S.~I.}\ \bibnamefont {Vilchinskii}},\
  }\href {\doibase 10.1103/PhysRevA.91.033607} {\bibfield  {journal} {\bibinfo
  {journal} {Phys. Rev. A}\ }\textbf {\bibinfo {volume} {91}},\ \bibinfo
  {pages} {033607} (\bibinfo {year} {2015})}\BibitemShut {NoStop}%
\bibitem [{\citenamefont {Hallwood}\ \emph {et~al.}(2006)\citenamefont
  {Hallwood}, \citenamefont {Burnett},\ and\ \citenamefont
  {Dunningham}}]{hallwood2006macroscopic}%
  \BibitemOpen
  \bibfield  {author} {\bibinfo {author} {\bibfnamefont {D.~W.}\ \bibnamefont
  {Hallwood}}, \bibinfo {author} {\bibfnamefont {K.}~\bibnamefont {Burnett}}, \
  and\ \bibinfo {author} {\bibfnamefont {J.}~\bibnamefont {Dunningham}},\
  }\href@noop {} {\bibfield  {journal} {\bibinfo  {journal} {New J. Phys.}\
  }\textbf {\bibinfo {volume} {8}},\ \bibinfo {pages} {180} (\bibinfo {year}
  {2006})}\BibitemShut {NoStop}%
\bibitem [{\citenamefont {Solenov}\ and\ \citenamefont
  {Mozyrsky}(2010)}]{solenov2010metastable}%
  \BibitemOpen
  \bibfield  {author} {\bibinfo {author} {\bibfnamefont {D.}~\bibnamefont
  {Solenov}}\ and\ \bibinfo {author} {\bibfnamefont {D.}~\bibnamefont
  {Mozyrsky}},\ }\href@noop {} {\bibfield  {journal} {\bibinfo  {journal}
  {Phys. Rev. Lett.}\ }\textbf {\bibinfo {volume} {104}},\ \bibinfo {pages}
  {150405} (\bibinfo {year} {2010})}\BibitemShut {NoStop}%
\bibitem [{\citenamefont {Schenke}\ \emph {et~al.}(2011)\citenamefont
  {Schenke}, \citenamefont {Minguzzi},\ and\ \citenamefont
  {Hekking}}]{schenke2011nonadiabatic}%
  \BibitemOpen
  \bibfield  {author} {\bibinfo {author} {\bibfnamefont {C.}~\bibnamefont
  {Schenke}}, \bibinfo {author} {\bibfnamefont {A.}~\bibnamefont {Minguzzi}}, \
  and\ \bibinfo {author} {\bibfnamefont {F.}~\bibnamefont {Hekking}},\
  }\href@noop {} {\bibfield  {journal} {\bibinfo  {journal} {Physical Review
  A}\ }\textbf {\bibinfo {volume} {84}},\ \bibinfo {pages} {053636} (\bibinfo
  {year} {2011})}\BibitemShut {NoStop}%
\bibitem [{\citenamefont {Amico}\ \emph {et~al.}(2014)\citenamefont {Amico},
  \citenamefont {Aghamalyan}, \citenamefont {Auksztol}, \citenamefont {Crepaz},
  \citenamefont {Dumke},\ and\ \citenamefont {Kwek}}]{amico2014superfluid}%
  \BibitemOpen
  \bibfield  {author} {\bibinfo {author} {\bibfnamefont {L.}~\bibnamefont
  {Amico}}, \bibinfo {author} {\bibfnamefont {D.}~\bibnamefont {Aghamalyan}},
  \bibinfo {author} {\bibfnamefont {F.}~\bibnamefont {Auksztol}}, \bibinfo
  {author} {\bibfnamefont {H.}~\bibnamefont {Crepaz}}, \bibinfo {author}
  {\bibfnamefont {R.}~\bibnamefont {Dumke}}, \ and\ \bibinfo {author}
  {\bibfnamefont {L.~C.}\ \bibnamefont {Kwek}},\ }\href@noop {} {\bibfield
  {journal} {\bibinfo  {journal} {Sci. Rep.}\ }\textbf {\bibinfo {volume} {4}}
  (\bibinfo {year} {2014})}\BibitemShut {NoStop}%
\bibitem [{\citenamefont {Aghamalyan}\ \emph {et~al.}(2015)\citenamefont
  {Aghamalyan}, \citenamefont {Cominotti}, \citenamefont {Rizzi}, \citenamefont
  {Rossini}, \citenamefont {Hekking}, \citenamefont {Minguzzi}, \citenamefont
  {Kwek},\ and\ \citenamefont {Amico}}]{aghamalyan2015coherent}%
  \BibitemOpen
  \bibfield  {author} {\bibinfo {author} {\bibfnamefont {D.}~\bibnamefont
  {Aghamalyan}}, \bibinfo {author} {\bibfnamefont {M.}~\bibnamefont
  {Cominotti}}, \bibinfo {author} {\bibfnamefont {M.}~\bibnamefont {Rizzi}},
  \bibinfo {author} {\bibfnamefont {D.}~\bibnamefont {Rossini}}, \bibinfo
  {author} {\bibfnamefont {F.}~\bibnamefont {Hekking}}, \bibinfo {author}
  {\bibfnamefont {A.}~\bibnamefont {Minguzzi}}, \bibinfo {author}
  {\bibfnamefont {L.~C.}\ \bibnamefont {Kwek}}, \ and\ \bibinfo {author}
  {\bibfnamefont {L.}~\bibnamefont {Amico}},\ }\href@noop {} {\bibfield
  {journal} {\bibinfo  {journal} {New J. Phys.}\ }\textbf {\bibinfo {volume}
  {17}},\ \bibinfo {pages} {045023} (\bibinfo {year} {2015})}\BibitemShut
  {NoStop}%
\bibitem [{\citenamefont {Aghamalyan}\ \emph {et~al.}(2016)\citenamefont
  {Aghamalyan}, \citenamefont {Nguyen}, \citenamefont {Auksztol}, \citenamefont
  {Gan}, \citenamefont {Valado}, \citenamefont {Condylis}, \citenamefont
  {Kwek}, \citenamefont {Dumke},\ and\ \citenamefont
  {Amico}}]{aghamalyan2016atomtronic}%
  \BibitemOpen
  \bibfield  {author} {\bibinfo {author} {\bibfnamefont {D.}~\bibnamefont
  {Aghamalyan}}, \bibinfo {author} {\bibfnamefont {N.}~\bibnamefont {Nguyen}},
  \bibinfo {author} {\bibfnamefont {F.}~\bibnamefont {Auksztol}}, \bibinfo
  {author} {\bibfnamefont {K.}~\bibnamefont {Gan}}, \bibinfo {author}
  {\bibfnamefont {M.~M.}\ \bibnamefont {Valado}}, \bibinfo {author}
  {\bibfnamefont {P.}~\bibnamefont {Condylis}}, \bibinfo {author}
  {\bibfnamefont {L.}~\bibnamefont {Kwek}}, \bibinfo {author} {\bibfnamefont
  {R.}~\bibnamefont {Dumke}}, \ and\ \bibinfo {author} {\bibfnamefont
  {L.}~\bibnamefont {Amico}},\ }\href@noop {} {\bibfield  {journal} {\bibinfo
  {journal} {New J. Phys.}\ }\textbf {\bibinfo {volume} {18}},\ \bibinfo
  {pages} {075013} (\bibinfo {year} {2016})}\BibitemShut {NoStop}%
\bibitem [{\citenamefont {Mathey}\ and\ \citenamefont
  {Mathey}(2016)}]{Mathey_Mathey2016}%
  \BibitemOpen
  \bibfield  {author} {\bibinfo {author} {\bibfnamefont {A.~C.}\ \bibnamefont
  {Mathey}}\ and\ \bibinfo {author} {\bibfnamefont {L.}~\bibnamefont
  {Mathey}},\ }\href {http://stacks.iop.org/1367-2630/18/i=5/a=055016}
  {\bibfield  {journal} {\bibinfo  {journal} {New J. Phys.}\ }\textbf {\bibinfo
  {volume} {18}},\ \bibinfo {pages} {055016} (\bibinfo {year}
  {2016})}\BibitemShut {NoStop}%
\bibitem [{\citenamefont {Haug}\ \emph {et~al.}(2017)\citenamefont {Haug},
  \citenamefont {Heimonen}, \citenamefont {Dumke}, \citenamefont {Kwek},\ and\
  \citenamefont {Amico}}]{haug2017aharonov}%
  \BibitemOpen
  \bibfield  {author} {\bibinfo {author} {\bibfnamefont {T.}~\bibnamefont
  {Haug}}, \bibinfo {author} {\bibfnamefont {H.}~\bibnamefont {Heimonen}},
  \bibinfo {author} {\bibfnamefont {R.}~\bibnamefont {Dumke}}, \bibinfo
  {author} {\bibfnamefont {L.-C.}\ \bibnamefont {Kwek}}, \ and\ \bibinfo
  {author} {\bibfnamefont {L.}~\bibnamefont {Amico}},\ }\href@noop {}
  {\bibfield  {journal} {\bibinfo  {journal} {arXiv:1706.05180}\ } (\bibinfo
  {year} {2017})}\BibitemShut {NoStop}%
\bibitem [{\citenamefont {Haug}\ \emph
  {et~al.}(2018{\natexlab{a}})\citenamefont {Haug}, \citenamefont {Tan},
  \citenamefont {Theng}, \citenamefont {Dumke}, \citenamefont {Kwek},\ and\
  \citenamefont {Amico}}]{haug2018readout}%
  \BibitemOpen
  \bibfield  {author} {\bibinfo {author} {\bibfnamefont {T.}~\bibnamefont
  {Haug}}, \bibinfo {author} {\bibfnamefont {J.}~\bibnamefont {Tan}}, \bibinfo
  {author} {\bibfnamefont {M.}~\bibnamefont {Theng}}, \bibinfo {author}
  {\bibfnamefont {R.}~\bibnamefont {Dumke}}, \bibinfo {author} {\bibfnamefont
  {L.-C.}\ \bibnamefont {Kwek}}, \ and\ \bibinfo {author} {\bibfnamefont
  {L.}~\bibnamefont {Amico}},\ }\href@noop {} {\bibfield  {journal} {\bibinfo
  {journal} {Phys. Rev. A}\ }\textbf {\bibinfo {volume} {97}},\ \bibinfo
  {pages} {013633} (\bibinfo {year} {2018}{\natexlab{a}})}\BibitemShut
  {NoStop}%
\bibitem [{\citenamefont {Haug}\ \emph
  {et~al.}(2018{\natexlab{b}})\citenamefont {Haug}, \citenamefont {Dumke},
  \citenamefont {Kwek},\ and\ \citenamefont {Amico}}]{haug2018andreev}%
  \BibitemOpen
  \bibfield  {author} {\bibinfo {author} {\bibfnamefont {T.}~\bibnamefont
  {Haug}}, \bibinfo {author} {\bibfnamefont {R.}~\bibnamefont {Dumke}},
  \bibinfo {author} {\bibfnamefont {L.-C.}\ \bibnamefont {Kwek}}, \ and\
  \bibinfo {author} {\bibfnamefont {L.}~\bibnamefont {Amico}},\ }\href@noop {}
  {\bibfield  {journal} {\bibinfo  {journal} {arXiv:1807.03616}\ } (\bibinfo
  {year} {2018}{\natexlab{b}})}\BibitemShut {NoStop}%
\bibitem [{\citenamefont {Seaman}\ \emph {et~al.}(2007)\citenamefont {Seaman},
  \citenamefont {Kr{\"a}mer}, \citenamefont {Anderson},\ and\ \citenamefont
  {Holland}}]{seaman2007atomtronics}%
  \BibitemOpen
  \bibfield  {author} {\bibinfo {author} {\bibfnamefont {B.}~\bibnamefont
  {Seaman}}, \bibinfo {author} {\bibfnamefont {M.}~\bibnamefont {Kr{\"a}mer}},
  \bibinfo {author} {\bibfnamefont {D.}~\bibnamefont {Anderson}}, \ and\
  \bibinfo {author} {\bibfnamefont {M.}~\bibnamefont {Holland}},\ }\href@noop
  {} {\bibfield  {journal} {\bibinfo  {journal} {Phys. Rev. A}\ }\textbf
  {\bibinfo {volume} {75}},\ \bibinfo {pages} {023615} (\bibinfo {year}
  {2007})}\BibitemShut {NoStop}%
\bibitem [{\citenamefont {Amico}\ \emph {et~al.}(2005)\citenamefont {Amico},
  \citenamefont {Osterloh},\ and\ \citenamefont
  {Cataliotti}}]{amico2005quantum}%
  \BibitemOpen
  \bibfield  {author} {\bibinfo {author} {\bibfnamefont {L.}~\bibnamefont
  {Amico}}, \bibinfo {author} {\bibfnamefont {A.}~\bibnamefont {Osterloh}}, \
  and\ \bibinfo {author} {\bibfnamefont {F.}~\bibnamefont {Cataliotti}},\
  }\href@noop {} {\bibfield  {journal} {\bibinfo  {journal} {Phys. Rev. Lett.}\
  }\textbf {\bibinfo {volume} {95}},\ \bibinfo {pages} {063201} (\bibinfo
  {year} {2005})}\BibitemShut {NoStop}%
\bibitem [{\citenamefont {Dumke}\ \emph {et~al.}(2016)\citenamefont {Dumke},
  \citenamefont {Lu}, \citenamefont {Close}, \citenamefont {Robins},
  \citenamefont {Weis}, \citenamefont {Mukherjee}, \citenamefont {Birkl},
  \citenamefont {Hufnagel}, \citenamefont {Amico}, \citenamefont {Boshier}
  \emph {et~al.}}]{dumke2016roadmap}%
  \BibitemOpen
  \bibfield  {author} {\bibinfo {author} {\bibfnamefont {R.}~\bibnamefont
  {Dumke}}, \bibinfo {author} {\bibfnamefont {Z.}~\bibnamefont {Lu}}, \bibinfo
  {author} {\bibfnamefont {J.}~\bibnamefont {Close}}, \bibinfo {author}
  {\bibfnamefont {N.}~\bibnamefont {Robins}}, \bibinfo {author} {\bibfnamefont
  {A.}~\bibnamefont {Weis}}, \bibinfo {author} {\bibfnamefont {M.}~\bibnamefont
  {Mukherjee}}, \bibinfo {author} {\bibfnamefont {G.}~\bibnamefont {Birkl}},
  \bibinfo {author} {\bibfnamefont {C.}~\bibnamefont {Hufnagel}}, \bibinfo
  {author} {\bibfnamefont {L.}~\bibnamefont {Amico}}, \bibinfo {author}
  {\bibfnamefont {M.~G.}\ \bibnamefont {Boshier}},  \emph {et~al.},\
  }\href@noop {} {\bibfield  {journal} {\bibinfo  {journal} {Journal of
  Optics}\ }\textbf {\bibinfo {volume} {18}},\ \bibinfo {pages} {093001}
  (\bibinfo {year} {2016})}\BibitemShut {NoStop}%
\bibitem [{\citenamefont {Amico}\ \emph {et~al.}(2017)\citenamefont {Amico},
  \citenamefont {Birkl}, \citenamefont {Boshier},\ and\ \citenamefont
  {Kwek}}]{Amico_NJP}%
  \BibitemOpen
  \bibfield  {author} {\bibinfo {author} {\bibfnamefont {L.}~\bibnamefont
  {Amico}}, \bibinfo {author} {\bibfnamefont {G.}~\bibnamefont {Birkl}},
  \bibinfo {author} {\bibfnamefont {M.}~\bibnamefont {Boshier}}, \ and\
  \bibinfo {author} {\bibfnamefont {L.-C.}\ \bibnamefont {Kwek}},\ }\href
  {http://stacks.iop.org/1367-2630/19/i=2/a=020201} {\bibfield  {journal}
  {\bibinfo  {journal} {New J. Phys.}\ }\textbf {\bibinfo {volume} {19}},\
  \bibinfo {pages} {020201} (\bibinfo {year} {2017})}\BibitemShut {NoStop}%
\bibitem [{\citenamefont {Aghamalyan}\ \emph {et~al.}(2013)\citenamefont
  {Aghamalyan}, \citenamefont {Amico},\ and\ \citenamefont
  {Kwek}}]{aghamalyan2013effective}%
  \BibitemOpen
  \bibfield  {author} {\bibinfo {author} {\bibfnamefont {D.}~\bibnamefont
  {Aghamalyan}}, \bibinfo {author} {\bibfnamefont {L.}~\bibnamefont {Amico}}, \
  and\ \bibinfo {author} {\bibfnamefont {L.~C.}\ \bibnamefont {Kwek}},\
  }\href@noop {} {\bibfield  {journal} {\bibinfo  {journal} {Phys. Rev. A}\
  }\textbf {\bibinfo {volume} {88}},\ \bibinfo {pages} {063627} (\bibinfo
  {year} {2013})}\BibitemShut {NoStop}%
\bibitem [{\citenamefont {Victorin}\ \emph {et~al.}(2018)\citenamefont
  {Victorin}, \citenamefont {Hekking},\ and\ \citenamefont
  {Minguzzi}}]{victorin2018bosonic}%
  \BibitemOpen
  \bibfield  {author} {\bibinfo {author} {\bibfnamefont {N.}~\bibnamefont
  {Victorin}}, \bibinfo {author} {\bibfnamefont {F.}~\bibnamefont {Hekking}}, \
  and\ \bibinfo {author} {\bibfnamefont {A.}~\bibnamefont {Minguzzi}},\
  }\href@noop {} {\bibfield  {journal} {\bibinfo  {journal} {arXiv preprint
  arXiv:1803.02718}\ } (\bibinfo {year} {2018})}\BibitemShut {NoStop}%
\bibitem [{\citenamefont {Haug}\ \emph
  {et~al.}(2018{\natexlab{c}})\citenamefont {Haug}, \citenamefont {Amico},
  \citenamefont {Dumke},\ and\ \citenamefont {Kwek}}]{haug2018mesoscopic}%
  \BibitemOpen
  \bibfield  {author} {\bibinfo {author} {\bibfnamefont {T.}~\bibnamefont
  {Haug}}, \bibinfo {author} {\bibfnamefont {L.}~\bibnamefont {Amico}},
  \bibinfo {author} {\bibfnamefont {R.}~\bibnamefont {Dumke}}, \ and\ \bibinfo
  {author} {\bibfnamefont {L.-C.}\ \bibnamefont {Kwek}},\ }\href@noop {}
  {\bibfield  {journal} {\bibinfo  {journal} {Quantum Science and Technology}\
  }\textbf {\bibinfo {volume} {3}},\ \bibinfo {pages} {035006} (\bibinfo {year}
  {2018}{\natexlab{c}})}\BibitemShut {NoStop}%
\bibitem [{\citenamefont {Penrose}\ and\ \citenamefont
  {Onsager}(1956)}]{PenroseOnsager}%
  \BibitemOpen
  \bibfield  {author} {\bibinfo {author} {\bibfnamefont {O.}~\bibnamefont
  {Penrose}}\ and\ \bibinfo {author} {\bibfnamefont {L.}~\bibnamefont
  {Onsager}},\ }\href {\doibase 10.1103/PhysRev.104.576} {\bibfield  {journal}
  {\bibinfo  {journal} {Phys. Rev.}\ }\textbf {\bibinfo {volume} {104}},\
  \bibinfo {pages} {576} (\bibinfo {year} {1956})}\BibitemShut {NoStop}%
\bibitem [{\citenamefont {{Nozi\`eres, P.}}\ and\ \citenamefont {{Saint James,
  D.}}(1982)}]{Noziere}%
  \BibitemOpen
  \bibfield  {author} {\bibinfo {author} {\bibnamefont {{Nozi\`eres, P.}}}\
  and\ \bibinfo {author} {\bibnamefont {{Saint James, D.}}},\ }\href {\doibase
  10.1051/jphys:019820043070113300} {\bibfield  {journal} {\bibinfo  {journal}
  {J. Phys. France}\ }\textbf {\bibinfo {volume} {43}},\ \bibinfo {pages}
  {1133} (\bibinfo {year} {1982})}\BibitemShut {NoStop}%
\bibitem [{\citenamefont {Mueller}\ \emph {et~al.}(2006)\citenamefont
  {Mueller}, \citenamefont {Ho}, \citenamefont {Ueda},\ and\ \citenamefont
  {Baym}}]{MuellerFrag}%
  \BibitemOpen
  \bibfield  {author} {\bibinfo {author} {\bibfnamefont {E.~J.}\ \bibnamefont
  {Mueller}}, \bibinfo {author} {\bibfnamefont {T.-L.}\ \bibnamefont {Ho}},
  \bibinfo {author} {\bibfnamefont {M.}~\bibnamefont {Ueda}}, \ and\ \bibinfo
  {author} {\bibfnamefont {G.}~\bibnamefont {Baym}},\ }\href {\doibase
  10.1103/PhysRevA.74.033612} {\bibfield  {journal} {\bibinfo  {journal} {Phys.
  Rev. A}\ }\textbf {\bibinfo {volume} {74}},\ \bibinfo {pages} {033612}
  (\bibinfo {year} {2006})}\BibitemShut {NoStop}%
\bibitem [{\citenamefont {Ho}\ and\ \citenamefont {Yip}(2000)}]{Spin1Frag}%
  \BibitemOpen
  \bibfield  {author} {\bibinfo {author} {\bibfnamefont {T.-L.}\ \bibnamefont
  {Ho}}\ and\ \bibinfo {author} {\bibfnamefont {S.~K.}\ \bibnamefont {Yip}},\
  }\href {\doibase 10.1103/PhysRevLett.84.4031} {\bibfield  {journal} {\bibinfo
   {journal} {Phys. Rev. Lett.}\ }\textbf {\bibinfo {volume} {84}},\ \bibinfo
  {pages} {4031} (\bibinfo {year} {2000})}\BibitemShut {NoStop}%
\bibitem [{\citenamefont {Wilkin}\ \emph {et~al.}(1998)\citenamefont {Wilkin},
  \citenamefont {Gunn},\ and\ \citenamefont {Smith}}]{RotFrag}%
  \BibitemOpen
  \bibfield  {author} {\bibinfo {author} {\bibfnamefont {N.~K.}\ \bibnamefont
  {Wilkin}}, \bibinfo {author} {\bibfnamefont {J.~M.~F.}\ \bibnamefont {Gunn}},
  \ and\ \bibinfo {author} {\bibfnamefont {R.~A.}\ \bibnamefont {Smith}},\
  }\href {\doibase 10.1103/PhysRevLett.80.2265} {\bibfield  {journal} {\bibinfo
   {journal} {Phys. Rev. Lett.}\ }\textbf {\bibinfo {volume} {80}},\ \bibinfo
  {pages} {2265} (\bibinfo {year} {1998})}\BibitemShut {NoStop}%
\bibitem [{\citenamefont {Kawasaki}\ and\ \citenamefont
  {Holzmann}(2017)}]{kawasaki2017finite}%
  \BibitemOpen
  \bibfield  {author} {\bibinfo {author} {\bibfnamefont {E.}~\bibnamefont
  {Kawasaki}}\ and\ \bibinfo {author} {\bibfnamefont {M.}~\bibnamefont
  {Holzmann}},\ }\href@noop {} {\bibfield  {journal} {\bibinfo  {journal}
  {Physical Review A}\ }\textbf {\bibinfo {volume} {95}},\ \bibinfo {pages}
  {051601} (\bibinfo {year} {2017})}\BibitemShut {NoStop}%
\bibitem [{\citenamefont {Kolovsky}(2017)}]{kolovsky2017bogoliubov}%
  \BibitemOpen
  \bibfield  {author} {\bibinfo {author} {\bibfnamefont {A.~R.}\ \bibnamefont
  {Kolovsky}},\ }\href@noop {} {\bibfield  {journal} {\bibinfo  {journal}
  {Physical Review A}\ }\textbf {\bibinfo {volume} {95}},\ \bibinfo {pages}
  {033622} (\bibinfo {year} {2017})}\BibitemShut {NoStop}%
\bibitem [{Note1()}]{Note1}%
  \BibitemOpen
  \bibinfo {note} {The consequences of other choices for $\phi _A$ and $\phi
  _B$ are discussed in \cite {victorin2018bosonic}.}\BibitemShut {Stop}%
\bibitem [{\citenamefont {Jordan}\ and\ \citenamefont
  {Wigner}(1928)}]{Jordan1928}%
  \BibitemOpen
  \bibfield  {author} {\bibinfo {author} {\bibfnamefont {P.}~\bibnamefont
  {Jordan}}\ and\ \bibinfo {author} {\bibfnamefont {E.}~\bibnamefont
  {Wigner}},\ }\href {\doibase 10.1007/BF01331938} {\bibfield  {journal}
  {\bibinfo  {journal} {Zeitschrift f{\"u}r Physik}\ }\textbf {\bibinfo
  {volume} {47}},\ \bibinfo {pages} {631} (\bibinfo {year} {1928})}\BibitemShut
  {NoStop}%
\bibitem [{\citenamefont {Rigol}\ and\ \citenamefont
  {Muramatsu}(2005)}]{RigolHCB}%
  \BibitemOpen
  \bibfield  {author} {\bibinfo {author} {\bibfnamefont {M.}~\bibnamefont
  {Rigol}}\ and\ \bibinfo {author} {\bibfnamefont {A.}~\bibnamefont
  {Muramatsu}},\ }\href {\doibase 10.1103/PhysRevA.72.013604} {\bibfield
  {journal} {\bibinfo  {journal} {Phys. Rev. A}\ }\textbf {\bibinfo {volume}
  {72}},\ \bibinfo {pages} {013604} (\bibinfo {year} {2005})}\BibitemShut
  {NoStop}%
\bibitem [{\citenamefont {Li}\ \emph {et~al.}(2012)\citenamefont {Li},
  \citenamefont {Pitaevskii},\ and\ \citenamefont {Stringari}}]{Stringari-Li}%
  \BibitemOpen
  \bibfield  {author} {\bibinfo {author} {\bibfnamefont {Y.}~\bibnamefont
  {Li}}, \bibinfo {author} {\bibfnamefont {L.~P.}\ \bibnamefont {Pitaevskii}},
  \ and\ \bibinfo {author} {\bibfnamefont {S.}~\bibnamefont {Stringari}},\
  }\href {\doibase 10.1103/PhysRevLett.108.225301} {\bibfield  {journal}
  {\bibinfo  {journal} {Phys. Rev. Lett.}\ }\textbf {\bibinfo {volume} {108}},\
  \bibinfo {pages} {225301} (\bibinfo {year} {2012})}\BibitemShut {NoStop}%
\bibitem [{\citenamefont {Cha}\ and\ \citenamefont {Shin}(2011)}]{ChaMin2011}%
  \BibitemOpen
  \bibfield  {author} {\bibinfo {author} {\bibfnamefont {M.-C.}\ \bibnamefont
  {Cha}}\ and\ \bibinfo {author} {\bibfnamefont {J.-G.}\ \bibnamefont {Shin}},\
  }\href {\doibase 10.1103/PhysRevA.83.055602} {\bibfield  {journal} {\bibinfo
  {journal} {Phys. Rev. A}\ }\textbf {\bibinfo {volume} {83}},\ \bibinfo
  {pages} {055602} (\bibinfo {year} {2011})}\BibitemShut {NoStop}%
\end{thebibliography}%

\appendix

\section{Momentum distribution width}
We provide here some complementary material.

\begin{figure}[htbp]
	\centering
	\subfigimg[width=0.4\textwidth]{}{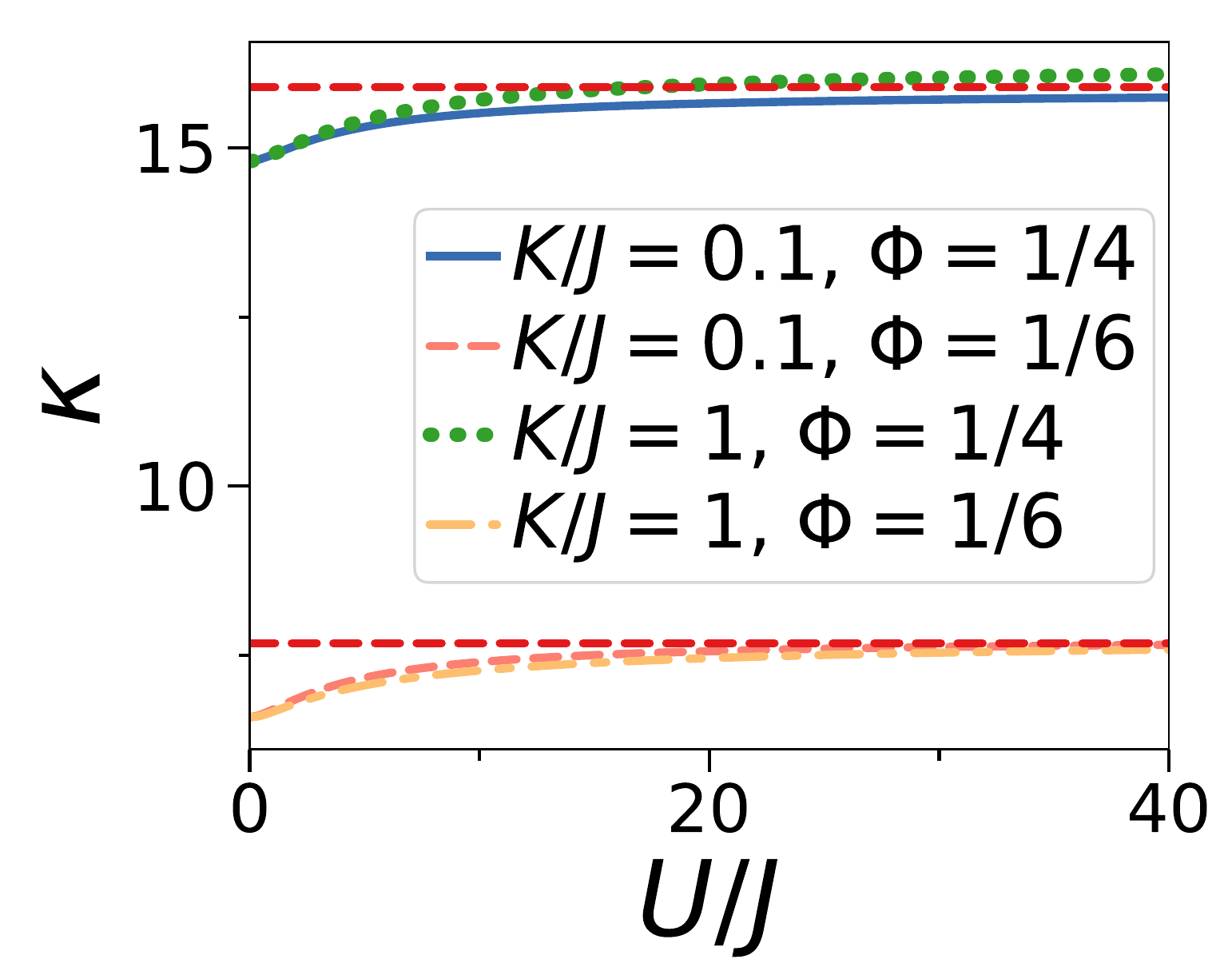}
	\caption{(Color online) Width of the momentum distribution $\kappa=\sum_k k^2 n_k$ as a function of interaction strength $U/J$ for various values of $\phi$ and $K$. The red dashed line is the corresponding value for non-interaction fermions. The calculations are performed for $N=6$ particles, $L=12$ sites.}
	\label{widthk}
\end{figure}

The width of the momentum distribution $\kappa=\sum_k k^2 n_k$ is shown in Fig.\ref{widthk}. At increasing interactions and small-interring coupling it tends to the fermionic value, thereby further confirming the fermionized nature of the state.

\end{document}